\documentclass[11pt]{article}

\usepackage{enumitem}
\usepackage{multirow}
\usepackage{chet}
\usepackage{graphics}
\usepackage{setspace}

\allowdisplaybreaks

\newcommand{\overbar}[1]{\mkern 3.5mu\overline{\mkern-3.5mu#1%
  \mkern-1.5mu}\mkern 1.5mu}
\newcommand{\overbarTheta}[1]{\mkern 1.5mu\overline{\mkern-1.5mu#1%
  \mkern-2mu}\mkern 2mu}

\newcommand\scalemath[2]{\scalebox{#1}[1]{\mbox{\ensuremath{\displaystyle #2}}}}

\newcommand{\CN}{\mathcal{N}}
\newcommand{\CO}{\mathcal{O}}
\newcommand{\COb}{{\overbar{\mathcal{O}}}}
\newcommand{\CJ}{\mathcal{J}}
\newcommand{\CT}{\mathcal{T}}
\newcommand{\CS}{\mathcal{S}}
\newcommand{\COind}[1]{\CO_{#1_1\hspace{-0.8pt}\ldots#1_j;\lsp
  \dot{#1}_1\hspace{-0.8pt}\ldots\dot{#1}_\jb}}
\newcommand{\COindb}[1]{\COb_{#1_1\hspace{-0.8pt}\ldots#1_j;\lsp
  \dot{#1}_1\hspace{-0.8pt}\ldots\dot{#1}_\jb}}

\newcommand{\COindellpI}[1]{\CO_{#1_1\hspace{-0.8pt}\ldots#1_{\ell+2};\lsp
  \dot{#1}_1\hspace{-0.8pt}\ldots\dot{#1}_{\ell}}}

\newcommand{\Qb}{{\overbar{Q}}{}}
\newcommand{\Db}{{\overbar{D}}{}}
\newcommand{\qb}{{\bar{q}}}

\newcommand{\jb}{{\bar{\jmath}}\hspace{0.9pt}}
\newcommand{\thetab}{{\bar{\theta}}}
\newcommand{\Thetab}{{\overbarTheta{\Theta}}{}}
\newcommand{\etap}{\eta^\prime}
\newcommand{\etab}{{\bar{\eta}}}
\newcommand{\etapb}{{\bar{\eta}^\prime}}

\newcommand{\sigmab}{{\bar{\sigma}}}
\newcommand{\Xb}{{\overbar{X}}{}}
\newcommand{\alphad}{{\dot{\alpha}}}
\newcommand{\alphapd}{{\dot{\alpha}^\prime}\hspace{-1pt}}
\newcommand{\alphap}{{\alpha^\prime}\hspace{-1pt}}
\newcommand{\betad}{{\smash{\dot{\beta}}}}
\newcommand{\betapd}{{\smash{\dot{\beta}^\prime}}\hspace{-1pt}}
\newcommand{\betap}{{\smash{\beta^\prime}}\hspace{-1pt}}
\newcommand{\gammad}{{\dot{\gamma}}}
\newcommand{\xup}{{\text{x}}}
\newcommand{\Uup}{{\text{U}}}
\newcommand{\vev}[1]{{\langle #1\rangle}}

\newcommand{\lsp}{\hspace{1pt}}
\newcommand{\llsp}{\hspace{0.5pt}}
\newcommand{\lnsp}{\hspace{-0.8pt}}

\newcommand{\AevenA}{{A_1}}
\newcommand{\AevenB}{{B_1}}
\newcommand{\AevenC}{{A_3}}
\newcommand{\AevenD}{{B_3}}
\newcommand{\AevenE}{{C_1}}
\newcommand{\AevenF}{{A_2}}
\newcommand{\AevenG}{{B_2}}
\newcommand{\AevenH}{{A_4}}
\newcommand{\AevenI}{{B_4}}
\newcommand{\AevenJ}{{C_2}}
\newcommand{\AevenK}{{C_3}}
\newcommand{\AevenL}{{C_4}}
\newcommand{\AevenM}{{C_5}}
\newcommand{\AevenN}{{C_6}}
\newcommand{\AevenO}{{C_7}}
\newcommand{\AevenP}{{C_8}}

\newcommand{\AoddA}{{A_1}}
\newcommand{\AoddB}{{B_1}}
\newcommand{\AoddC}{{A_2}}
\newcommand{\AoddD}{{B_2}}
\newcommand{\AoddE}{{C_3}}
\newcommand{\AoddF}{{D_3}}
\newcommand{\AoddG}{{E_1}}
\newcommand{\AoddH}{{E_2}}
\newcommand{\AoddI}{{C_2}}
\newcommand{\AoddJ}{{D_2}}
\newcommand{\AoddK}{{E_3}}
\newcommand{\AoddL}{{E_4}}
\newcommand{\AoddM}{{C_1}}
\newcommand{\AoddN}{{D_1}}
\newcommand{\AoddO}{{E_5}}
\newcommand{\AoddP}{{E_6}}

\newcommand{\BevenA}{{A_2}}
\newcommand{\BevenB}{{B_2}}
\newcommand{\BevenC}{{A_1}}
\newcommand{\BevenD}{{B_1}}
\newcommand{\BevenE}{{C_1}}
\newcommand{\BevenF}{{C_2}}
\newcommand{\BevenG}{{C_3}}
\newcommand{\BevenH}{{C_4}}
\newcommand{\BevenI}{{C_5}}
\newcommand{\BevenJ}{{C_6}}

\newcommand{\BoddA}{{A_1}}
\newcommand{\BoddB}{{B_1}}
\newcommand{\BoddC}{{A_2}}
\newcommand{\BoddD}{{B_2}}
\newcommand{\BoddE}{{C_7}}
\newcommand{\BoddF}{{C_2}}
\newcommand{\BoddG}{{C_3}}
\newcommand{\BoddH}{{C_4}}
\newcommand{\BoddI}{{C_5}}
\newcommand{\BoddJ}{{C_6}}
\newcommand{\BoddK}{{C_1}}
\newcommand{\BoddL}{{C_8}}
\newcommand{\BoddM}{{C_9}}

\newcommand{\CevenA}{{A}}
\newcommand{\CevenB}{{B}}
\newcommand{\CevenC}{{C_1}}
\newcommand{\CevenD}{{C_2}}

\newcommand{\CoddA}{{C_1}}
\newcommand{\CoddB}{{C_2}}
\newcommand{\CoddC}{{C_3}}
\newcommand{\CoddD}{{C_4}}

\usepackage{amssymb}
\newcommand{\TAEvenLorentz}[1]{\mathcal{P}^{(#1)}}
\newcommand{\TAOddLorentz}[2][]{\mathcal{P}^{(#2){#1}}}
\newcommand{\TAEven}[2][]{\mathcal{P}^{#1(#2)}_{\text{A}}}
\newcommand{\TAOdd}[2][]{\tilde{\mathcal{P}}^{#1(#2)}_{\text{A}}}
\newcommand{\TBEven}[2][]{\mathcal{P}^{#1(#2)}_{\text{B}}}
\newcommand{\TBOdd}[2][]{\tilde{\mathcal{P}}^{#1(#2)}_{\text{B}}}
\newcommand{\TCEven}[2][]{\mathcal{P}^{#1(#2)}_{\text{C}}}
\newcommand{\TCOdd}[2][]{\tilde{\mathcal{P}}^{#1(#2)}_{\text{C}}}

\newcommand{\Uc}[2]{\lsp [#1\ifthenelse{\equal{#2}{\Theta}}{\Thetab}{\bar{#2}}]}
\newcommand{\Ec}[2]{\lsp [#1#2]}
\newcommand{\Ecb}[2]{\lsp [\ifthenelse{\equal{#1}{\Theta}}{\Thetab}{\bar{#1}}
                           \ifthenelse{\equal{#2}{\Theta}}{\Thetab}{\bar{#2}}]}

\date{April 2018}

\preprint{CERN-TH-2018-096}

\title{R-current three-point functions\\\vspace{5pt}
in 4d $\mathcal{N}=1$ superconformal theories}

\author{Andrea Manenti,${\!}^a$ Andreas Stergiou,${\!}^b$
and Alessandro Vichi${}^a$}

\affiliation{${}^a$Institute of Physics, \'Ecole Polytechnique
F\'ed\'erale de Lausanne (EPFL),\\\vspace{-2pt}
Rte de la Sorge, BSP 728, CH-1015 Lausanne, Switzerland\\\vspace{2pt}
${}^b$Theoretical Physics Department, CERN, 1211 Geneva 23, Switzerland}

\abstract{In 4d $\mathcal N=1$ superconformal field theories (SCFTs) the
R-symmetry current, the stress-energy tensor, and the supersymmetry
currents are grouped into a single object, the Ferrara--Zumino multiplet.
In this work we study the most general form of three-point functions
involving two Ferrara--Zumino multiplets and a third generic multiplet. We
solve the constraints imposed by conservation in superspace and show that
non-trivial solutions can only be found if the third multiplet is R-neutral
and transforms in suitable Lorentz representations. In the process we give
a prescription for counting independent tensor structures in superconformal
three-point functions. Finally, we set the Grassmann coordinates of the
Ferrara--Zumino multiplets to zero and extract all three-point functions
involving two R-currents and a third conformal primary. Our results pave
the way for bootstrapping the correlation function of four R-currents in 4d
$\mathcal N=1$ SCFTs.}

\begin{document}

\maketitle

{\setstretch{1.12}
\toc}

\newsec{Introduction}
In the last decade the conformal bootstrap has been widely used to explore
the space of conformal field theories (CFTs), both from a numerical
perspective~\cite{Rattazzi:2008pe} but also from an analytical
one~\cite{Fitzpatrick:2012yx, Komargodski:2012ek}.  Important results have
been obtained for 3d condensed matter systems, but also for
higher-dimensional theories in presence of supersymmetry (see
\cite{Poland:2016chs} for a review and a summary of important results).

Most of these techniques heavily rely on the computation of conformal
blocks, or superconformal blocks in the case of superconformal field
theories (SCFTs).  While for CFTs this problem has now been completely
solved in 3d~\cite{Costa:2011dw,Iliesiu:2015akf} and
4d~\cite{Echeverri:2015rwa, Echeverri:2016dun}, it is still an
open question in several supersymmetric cases. Superconformal blocks can be
expressed as finite linear combinations of ordinary blocks. Nevertheless,
finding the exact expression is technically challenging.  A notable example
are the superconformal blocks of the stress-energy tensor multiplet four
point function in 4d $\mathcal N=2$ theories, see
e.g.~\cite{Liendo:2015ofa}.

In this work we focus on 4d $\CN=1$ theories. Computations of
superconformal blocks have already been performed in the literature for
four-point functions of scalar operators, e.g.\ chiral or
antichiral~\cite{Poland:2010wg, Vichi:2011ux}, linear~\cite{Fortin:2011nq,
Berkooz:2014yda}, and general~\cite{Khandker:2014mpa, Li:2016chh,
Li:2017ddj}.  Here we take the next logical step and address a more
complicated case, i.e.\ we compute all necessary ingredients for the
calculation of superconformal blocks of four-point functions involving the
R-current $J_\mu$. The vector operator $J_\mu$ is in the same multiplet as
the supersymmetry current and the stress-energy tensor, called the
Ferrara--Zumino multiplet, $\CJ_\mu$~\cite{Ferrara:1974pz}.  Our results
are obtained by explicitly working out the projection of the superconformal
three-point function with the Ferrara--Zumino multiplet at the first two
points and a general allowed operator at the third.  The main motivation is
to use these results to bootstrap the four-point function of $J_\mu$. This
will provide a new way to explore the space of SCFTs, and  hopefully shed
more light on the ``minimal'' 4d $\CN=1$ SCFT studied with bootstrap
techniques in~\cite{Poland:2011ey, Poland:2015mta, Li:2017ddj} and
attempted to be identified by analytical means in~\cite{Buican:2016hnq,
Xie:2016hny}.

Unlike the case of extended supersymmetry in 4d, in $\mathcal N=1$ SCFTs
the supermultiplet containing the stress-energy tensor does not contain a
scalar primary operator. Dealing with spinning operators raises the
complication of the computations significantly, although the general
procedure remains the same.

Let us outline the logic we follow here. Our starting point is the
superconformal three-point function in superspace with the Ferrara--Zumino
multiplet at the first two points and a general allowed superconformal
multiplet at the third.\foot{There are various types of allowed operators
at the third point, belonging to different Lorentz representations and
having appropriate R-charges.} The superspace expression for the
three-point function, constructed following the constraints laid out
in~\cite{Park:1997bq, Osborn:1998qu}, involves many structures with \emph{a
priori} independent coefficients. The first step is to work out the relations
among these coefficients due to the shortening condition
satisfied by the Ferrara--Zumino multiplet. Typically, this reduces the
number of independent coefficients drastically, and in some cases sets the
whole three-point function to zero. Subsequently, we perform an expansion
in the fermionic coordinates $\theta_3$ and $\thetab_3$, after setting
$\theta_1=\theta_2=0$ and $\thetab_1=\thetab_2=0$ in order to focus on the
R-current at the first two points. With this expansion we are able to
identify three-point functions of conformal primary operators.

This last step is the most complicated and delicate one: any given order of
the expansion contains a combination of conformal primaries and descendants
which must be disentangled.  For example, at order $\theta_3\thetab_3$ the
expansion of the superconformal three-point function contains not only
terms belonging to three-point functions of the schematic form $\langle
JJ(Q\Qb\CO)_p\rangle$, where $Q$ is the supersymmetric charge and ``$p$''
denotes that the operator is conformal primary, but also terms belonging to
three-point functions of the schematic form $\langle JJ(P\CO)\rangle$,
where $P$ is the generator of translations. The latter contributions can be
subtracted away using the results of~\cite{Li:2014gpa}, where the specific
way contamination from conformal descendants can happen was worked out in
generality. To carry out our calculations we have expanded the
\emph{Mathematica} package\footnote{The package can be made available upon
request.} developed for the purposes of~\cite{Li:2014gpa}.  For the
structures associated with three-point functions of conformal primary
operators we have used the \emph{Mathematica} package
\href{https://gitlab.com/bootstrapcollaboration/CFTs4D}{\texttt{CFTs4D}}~\cite{Cuomo:2017wme}.

A non-trivial check on our computations is supplied by the fact that when
the third operator in the three-point function satisfies a shortening
condition, then a unitarity bound is saturated and the corresponding
three-point function should vanish. This typically happens automatically
after the Ward identities for conservation at the first two points have
been solved, i.e.\ the solution for the independent three-point function
coefficients involves explicit factors of $\Delta-\Delta_u$, where
$\Delta_u$ is the dimension at the unitarity bound. While in some of our
cases this story is repeated, we have also encountered situations where
solving the Ward identities at the first two points is not enough to
guarantee vanishing of the three-point function when the third operator
saturates its unitarity bound. The Ward identity at the third point needs
to be imposed in those cases, something that results in the proper
vanishing of the three-point function. In some cases the resulting
requirement is non-trivial, i.e.\ it does not set all independent (after
satisfying the Ward identity at the first two points) coefficients to zero
at the unitarity bound, but rather it relates them in the appropriate way.

In Sec.~\ref{Warmup} we review known results about the structure of
three-point functions with two conserved spin-one currents at the first two
points. In Sec.~\ref{secThreePFA} we explore general constraints on our
three-point functions of interest in superspace. We introduce an index-free
notation and we provide counting arguments for the number of independent
structures in superconformal three-point functions with two Ferrara--Zumino
and a general multiplet in $\CN=1$ superspace. The analysis of the Ward
identity constraints is contained in Sec.~\ref{Ward}, while our final
results for the various three-point function coefficients can be found in
Sec.~\ref{ThetaThreeExpansion}.  Appendix~\ref{appendix:structureEven}
contains the superconformal three-point function structures in the various
cases of interest.  Appendix~\ref{appendix:structureLorentz} contains
structures using Lorentz vector indices for the case where the third
operator in the three-point function is an integer-spin representation of
the Lorentz group. Some special cases regarding the solutions of the Ward
identity constraints are treated in Appendix~\ref{appendix:WISpinZeroOne},
while conventions for the supersymmetric derivatives used for the
implementation of the Ward identities are included in
Appendix~\ref{Conventions}. A \emph{Mathematica} file with a summary of all
results in also attached to this submission.

\newsec{Warming up:~\texorpdfstring{$\boldsymbol{\langle J_\mu J_\nu
\CO_{\ell+k,\ell}\rangle}$}{<JJO>} three-point function in CFTs}
\label{Warmup}

Before diving into the complications of supersymmetric CFTs, let us briefly
review the structure of the three-point function involving two identical
conserved spin-one currents $J_\mu$ and a third conformal primary operator
$\CO_{\ell+k,\ell}$, with scaling dimension $\Delta$, in a generic CFT in
4d. The integers $\ell$ and $k$ determine the transformation properties of
$\CO_{\ell+k,\ell}$ under the Lorentz group $\text{SO}(1,3)$. The most
general correlation function consistent with conformal symmetry can be
compactly written in embedding space; following \cite{Elkhidir:2014woa} we
lift the fields to 6d and we introduce 6d spinors $S_{i\alpha}$ and
$\bar{S}_i^{\dot{\alpha}}$, $i=1,2,3$, to contract the spacetime indices.
The three-point function is non-zero in three cases:
\eqn{\langle J(X_1) J(X_2) \CO_{\ell+k,\ell}(X_3)\rangle = \mathcal
  K_{\Delta,\ell,k}
 \left\{
\begin{array}{lc}
\displaystyle\sum_{i=1}^5 \lambda^{(i)} \mathcal
S^{(i)}(X_i,S_i,\overbar{S}_i) + \lambda^{(-)} \mathcal
S^{(-)}(X_i,S_i,\overbar{S}_i)  &  \text{for }k=0\,,\\
\displaystyle\sum_{i=1}^4 \lambda^{(i)} \mathcal
T^{(i)}(X_i,S_i,\overbar{S}_i) & \text{for }k=2\,,\\
\lambda\lsp \mathcal R(X_i,S_i,\overbar{S}_i)   & \text{for }k=4\,,\\
\end{array}
\right.
}[JJOnonsusy]
where the prefactor is
\eqn{
\mathcal K_{\Delta,\ell,k}
=\frac{J_3^{\ell-2}}{X_{12}^{\Delta+\ell-8+\frac12
k}X_{13}^{\Delta-\ell-\frac12 k}X_{23}^{\Delta-\ell-\frac12
k}}\,,}[JJOnonsusyPrefactor]
and we have defined the tensor structures
\begin{align}
  \mathcal S^{(1)} &=  J_1J_2J_3^2\,,
  & \mathcal T^{(1)} &=  I_{13}I_{23}I_{32} K_2 J_3\,,     \notag \\
\mathcal S^{(2)} &=  I_{23}I_{32} J_1J_3\,,                            & \mathcal T^{(2)} &=   I_{13}I_{23}I_{31} K_1 J_3\,,     \notag \\
\mathcal S^{(3)} &=   I_{13}I_{31} J_2 J_3\,,           & \mathcal T^{(3)} &=  I_{13}I_{21}K_1 J_3^2\,,     \notag \\
\mathcal S^{(4)} &=   I_{13}I_{31}I_{23}I_{32}\,,           & \mathcal T^{(4)} &=  I_{12}I_{23} K_2 J_3^2\,,      \notag \\
\mathcal S^{(5)} &=   I_{12}I_{21} J_3^2\,,            & \mathcal R &=     I_{13}I_{23} K_1 K_2 J_3^2\,,             \notag \\
\mathcal S^{(-)} &=I_{12}I_{23}I_{31}+I_{13}I_{12}I_{32}\,.
\label{TensorStruct}
\end{align}
The quantities appearing in \JJOnonsusyPrefactor and (\ref{TensorStruct})
are combinations of the 6d coordinates $X_i$ and spinors $S_{i\alpha}$ and
$\bar{S}_i^{\dot{\alpha}}$. Their definitions can be found in
\cite{Elkhidir:2014woa}.\footnote{Schematically we have: $X_{ij}=X_i\cdot
X_j$, $I_{ij}=\overbar{S}_i S_j$, $J_{i,jk}=\overbar{S}_i \mathbf{X}_j
\overbarTheta{\mathbf{X}}_k S_i/X_{jk}$, $K_{i,jk}=S_j
\overbarTheta{\mathbf{X}}_i S_k \sqrt{X_{jk}/(X_{ij}X_{ik})}$. In addition,
$\mathbf{X}_{ab}$ is obtained contracting $X^M$ with the 6d gamma matrices.
 Finally $J_1\equiv J_{1,23}$, $J_2\equiv J_{2,13}$, $J_3\equiv J_{3,12}$
 and similarly for $K_i$. In
 \href{https://gitlab.com/bootstrapcollaboration/CFTs4D}{\texttt{CFTs4D}}
 the structures $I,J,K$ are denoted $\hat I, \hat J, \hat K$.}
Notice that for the special cases $\ell=0,1$, not all of the above tensor
structures are allowed.  The case of $\CO_{\ell,\ell+k}$ is similar to
\JJOnonsusy: one can define tensor structures $\overbar{\mathcal T}^{(i)}$
and $\overbar{\mathcal{R}}$ analogous to (\ref{TensorStruct}) with the
replacement $I_{ij}K_m \rightarrow I_{ji}\overbar{K}_m$.

Permutation symmetry and conservation of the current $J_\mu$ impose a  set
of conditions summarized in Table~\ref{TableConservation}.  We found two
special cases. First, when $k=0$, $\ell=0$, then $\lambda^{(4)}=0$ since
the associated tensor structure does not exist. As a consequence,
$\lambda^{(2)}$ vanishes as well and there is only one degree of freedom. A
second exception is for $k=2$, $\ell=0$; in this case permutation symmetry
and current conservation sets the three-point function to zero, expect for
the special case $\Delta=2$, when $\lambda^{(3)}=\lambda^{(4)}$, while all
the rest vanishes. However, in SCFTs this operator is below the unitarity
bounds (see Sec.~\ref{Ward}).  Besides these special cases, we stress that
all the denominators in Table~\ref{TableConservation} are non-zero whenever
the dimension of $\mathcal O_{\ell+k,\ell}$ satisfies the unitarity bounds
\cite{Ferrara:1974pt,Mack:1975je}
\eqna{\begin{aligned}
& \Delta\geq \ell+\tfrac12 k+2\, \qquad& \text{for }\ell>0\,,\\
& \Delta\geq \tfrac12 k+1 \qquad & \text{for }\ell=0\,.
\end{aligned}}[NonSusyUnitarity]

\begin{table}[h!]
\begin{center}
\renewcommand{\arraystretch}{1.3}
\small
\begin{tabular}{|c|c|c|c|c|}
\hline
$k$ & $\ell$ & $1\leftrightarrow 2$ & Conservation & d.o.f\\
\hline
0 & Even & $ \lambda^{(3)}=-\lambda^{(2)}$, $\lambda^{(-)}=0$ & $\sigma\lambda^{(2)} = 4\ell(\Delta-3)\lambda^{(1)} - (\Delta-\ell-4)(\Delta+\ell)\lambda^{(4)}$,& $2^+$\\
 &  & & $ \sigma\lambda^{(5)} = 4(\Delta-3)(\Delta-2)\lambda^{(1)} + (\Delta-\ell-4)(\Delta-\ell-6)\lambda^{(4)}$& \\
0 & Odd & $\lambda^{(3)}=\lambda^{(2)}$, $\lambda^{(1,4,5)}=0$ & $\lambda^{(2)} =0$ & $1^-$\\
\hline
 2 & Even & $\lambda^{(2)}=-\lambda^{(1)}$, $\lambda^{(4)}=\lambda^{(3)}$ & $\lambda^{(3)} =- \frac{\Delta-\ell-6}{2(\Delta-2)}\lambda^{(1)}$& 1\\
 2 & Odd &$\lambda^{(2)}=\lambda^{(1)}$, $\lambda^{(4)}=-\lambda^{(3)}$ & $\lambda^{(3)} =-\frac{\Delta-\ell-6}{2(\ell+2)}\lambda^{(1)}$& 1\\
\hline
4 & Even & -  & - & 1\\
4 & Odd &$\lambda =0$  & - & 0\\
\hline
\end{tabular}
\renewcommand{\arraystretch}{1}
\end{center}
\caption{Constraints imposed by permutation symmetry and current
conservation. The last column shows the number of independent coefficients
after all conditions are imposed. The case $\ell=0$ is special. Here we
defined $\sigma=2 \ell  (\ell +8)-4(\ell-1)\Delta-2\Delta^2$.  }
\label{TableConservation}
\end{table}
\normalsize

\newsec{Three-point function of two Ferrara--Zumino and a general
multiplet}[secThreePFA]
Let us now move on to the supersymmetric case. For our analysis we will mostly
follow the formalism developed in~\cite{Park:1997bq, Osborn:1998qu} and the conventions of Wess and Bagger~\cite{Wess:1992cp}. With
$\mathcal N=1$ supersymmetry the conserved currents arising from
superconformal transformations are contained in a single superconformal
multiplet, the Ferrara--Zumino multiplet $\CJ_{\alpha\alphad}$
\cite{Ferrara:1974pz}. This satisfies the shortening condition\footnote{See Appendix~\ref{Conventions} for the definitions of superspace derivatives $D^\alpha$ and $\overbar{D}^\alphad$.}
\eqn{D^\alpha\CJ_{\alpha\alphad}=\overbar{D}^\alphad\CJ_{\alpha\alphad}
=0\,.}[SupercurCons]
The expansion of $\CJ_{\alpha\alphad}$ in components reads (see, e.g.\ \cite{Komargodski:2010rb})
\eqna{
-\tfrac{1}{2}\lsp\sigmab^{\alphad\alpha}_\mu\llsp\CJ_{\alpha\alphad}(z) &= J_\mu(x) + \tfrac{i}{2}\llsp \theta^{\alpha} S_{\mu\alpha}(x) -\tfrac{i}{2} \llsp \overbar{S}_{\mu\alphad}(x) \llsp\thetab^{\alphad}
+ \theta^\alpha\sigma^\nu_{\alpha\alphad}\thetab^\alphad\lsp \left(T_{\mu\nu}(x) - \tfrac{1}{2} \epsilon_{\mu\nu\rho\lambda} \partial^{\rho}J^{\lambda}(x)\right)
\\
&
\quad-\tfrac{1}{8}\theta^2  \lsp \lsp \partial_\nu S_{\mu}(x)\sigma^{\nu}\thetab
-\tfrac{1}{8}\thetab^2\lsp \lsp \theta\sigma^{\nu}\partial_\nu\overbar{S}_{\mu}(x)
-\tfrac{1}{4}\theta^2\thetab^2\lsp \partial^2 J_\mu(x) \,, }[SuperfieldJ]
where $z=(x,\theta,\thetab)$ is a point in superspace, $J_\mu$ is the
R-symmetry $U(1)$ current, $S_{\mu\alpha}$ the supersymmetry current and
$T_{\mu\nu}$ the stress-energy tensor. The condition \SupercurCons implies
the following conservation and irreducibility conditions:
\eqn{
\partial_\mu J^\mu = \partial_\mu T^{\mu\nu} = T^\mu{\!}_\mu = T_{[\mu\nu]} = \partial_\mu S^\mu_\alpha = \partial_\mu \overbar{S}^\mu_\alphad = \sigmab_\mu^{\alphad\alpha}	S^\mu_\alpha = \overbar{S}^\mu_\alphad \,\sigmab_\mu^{\alphad\alpha} =0\,.
}[conservation]

\subsec{General properties}
\label{General}

In this section we study the most general form of the three-point function
of two Ferrara--Zumino multiplets and a third general superconformal
multiplet $\COind{\gamma}=\CO_{(\gamma_1\hspace{-0.8pt}\ldots\gamma_j);
\lsp(\gammad_1\hspace{-0.8pt}\ldots\gammad_\jb)}$ consistent with 4d
$\CN=1$ superconformal symmetry. We recall that superconformal multiplets
are labelled by two integers, $j$ and $\jb$, indicating that the
superconformal primary in the multiplet transforms in the $(\tfrac12
j,\tfrac12\jb)$ representation of the Lorentz group,\foot{In
Sec.~\ref{Warmup} we considered operators with $j=\ell+k$, $\jb=\ell$.} and
two reals, $q$ and $\qb$, which give the scaling dimension and R-charge of
the superconformal primary operator via
\eqn{\Delta=q+\qb\,,\qquad R=\tfrac23(q-\qb)\,.}[]
While the supercurrent satisfies $q_{\CJ}=\qb_{\CJ}=\frac32$, for a general supermultiplet
$\CO$ the values $q,\qb$ can assume any value consistent with the unitarity bounds\footnote{Chiral (antichiral) representations are special cases and correspond to $\qb=\jb=0$ ($q=j=0$).}~\cite{Flato:1983te, Dobrev:1985qv}:
\eqn{\Delta\geq\left|q-\qb-\tfrac12(j-\jb)\right|
  +\tfrac12(j+\jb)+2\,.}[GenUniBoundA]

Our goal is to start from the superspace expression of the three-point function
\eqn{\vev{\CJ_{\alpha\alphad}(z_1)\lsp \CJ_{\beta\betad}(z_2)\lsp
  \COind{\gamma}(z_3)}\,,}[ThreePointSusy]
and then solve the constraints imposed by the shortening condition
\footnote{In this work we use interchangeably the terminology ``shortening condition" and ``Ward identity".}
\SupercurCons.
From the results reviewed in Sec.~\ref{Warmup} we know that if we set
$\theta_i=\thetab_i=0$ in \ThreePointSusy, the only nonzero contributions come from superprimaries satisfying
\begin{enumerate}[label=\Alph{enumi}]
  \renewcommand{\labelenumi}{(\Alph{enumi})}
  \item\label{caseA} $R=0$, $j=\jb\,$,
  \item\label{caseB} $R=0$, $j=\jb+2\,$,
  \item\label{caseC} $R=0$, $j=\jb+4\,$,
\end{enumerate}
and their conjugates.  In superspace, however, we would expect the
correlator \ThreePointSusy to be non-vanishing  whenever there is a
non-zero three-point function between a component of $\mathcal O$ and any
pairs of the fields appearing in the expansion \SuperfieldJ. For instance,
when only $\theta_{1,2}=\thetab_{1,2}=0$, we could get a non-supersymmetric
three-point function between two currents $J$ and a superconformal
descendant, i.e.\ the representation corresponding to cases
\ref{caseA}--\ref{caseC} may only arise after the action of $Q$'s and/or
$\Qb$'s on some superconformal primary.\foot{Examples of this have appeared
before in the literature~\cite{Berkooz:2014yda, Li:2017ddj}.} Similarly,
since in CFTs the OPE of two stress-energy tensors $T_{\mu\nu}$ can contain
primaries with $j-\jb$ up to $\pm8$, we could expect that these cases
should be considered too.

One of the most important results of this work is showing that cases
\ref{caseA}--\ref{caseC}  are instead the only relevant ones: although it
is possible to construct other three-point functions in superspace, the
shortening condition \SupercurCons sets all of them to zero. In
Sec.~\ref{seccons} we give a group theoretic argument for this fact. In
addition, we have verified that the conservation conditions admit
non-trivial solutions only in the cases \ref{caseA}--\ref{caseC}. In this
paper we will present in detail only the non-vanishing cases.

Before analyzing each of the cases \ref{caseA}--\ref{caseC} individually,
let us discuss the general properties of the correlator \ThreePointSusy.
As shown in~\cite{Osborn:1998qu}, the most general three-point function
consistent with superconformal symmetry can be written as
\eqn{\vev{\CJ_{\alpha\alphad}(z_1)\lsp \CJ_{\beta\betad}(z_2)\lsp
  \COind{\gamma}(z_3)} =
  \frac{\xup_{1\bar{3}\lsp\alpha\alphapd}\lsp
  \xup_{3\bar{1}\lsp\alphap\alphad}\lsp
  \xup_{2\bar{3}\lsp\beta\betapd}\lsp
  \xup_{3\bar{2}\lsp\betap\betad}}
  {(x_{\bar{3}1}{\lnsp}^2\lsp x_{\bar{1}3}{\lnsp}^2\lsp
  x_{\bar{3}2}{\lnsp}^2\lsp x_{\bar{2}3}{\lnsp}^2)^2}
  \lsp t^{\alphapd\alphap;
  \lsp\betapd\betap}{\!}_{\gamma_1\hspace{-0.8pt}\ldots\gamma_j;
  \lsp\gammad_1\hspace{-0.8pt}\ldots\gammad_\jb}(X,\Theta,\Thetab)\,,
}[ThreePFAgen]
where\footnote{In \cite{Osborn:1998qu} $X,\,\Theta$ and $\Thetab$
correspond to, respectively, $X_3,\,\Theta_3$ and $\Thetab_3$.}
\eqn{\begin{gathered}
  {\text{X}}=\frac{{\xup}_{3\bar{1}}\tilde{\xup}_{\bar{1}
  2}{\xup}_{2\bar{3}}}{x_{\bar{1}3}{}^2 x_{\bar{3}2}{}^2}\,, \qquad
  \xup_{\alpha\alphad}=\sigma^\mu_{\alpha\alphad}x_\mu\,,\qquad
  \tilde{\xup}^{\alphad\alpha}=\epsilon^{\alpha\beta}
  \epsilon^{\alphad\betad}\xup_{\beta\betad}\,,\\
   \Theta=i\left(\frac{1}{x_{\bar{1}{3}}{\lnsp}^2}
  \xup_{3\bar{1}}\thetab_{31}
  -\frac{1}{x_{\bar{2}3}{\lnsp}^2}\xup_{3\bar{2}}\thetab_{32}\right),
  \qquad\Thetab=\Theta^\ast\,,
\end{gathered}}[Xdefn]
with $\thetab_{ij}=\thetab_i-\thetab_j$ and the supersymmetric interval
between $x_i$ and $x_j$ defined by
\eqn{x_{\bar{\imath}j}=-x_{j\bar{\imath}}\equiv x_{ij}
-i\llsp\theta_i\sigma\bar{\theta}_i
-i\llsp\theta_j\sigma\bar{\theta}_j
+2i\llsp\theta_j\sigma\bar{\theta}_i\,.
}[superdistance]
In addtion, the tensor $t$ must satisfy the homogeneity
property~\cite{Osborn:1998qu}
\eqn{t^{\alphad\alpha;
  \lsp\betad\beta}{\!}_{\gamma_1\hspace{-0.8pt}\ldots\gamma_j;
  \lsp\gammad_1\hspace{-0.8pt}\ldots\gammad_\jb}(\lambda\bar{\lambda}
  X,\lambda \Theta,\bar{\lambda} \Thetab) =
  \lambda^{\frac23(2q+\bar{q}-9)}\bar{\lambda}^{\frac23(q+2\bar{q}-9)} t^{\alphad\alpha;
  \lsp\betad\beta}{\!}_{\gamma_1\hspace{-0.8pt}\ldots\gamma_j;
  \lsp\gammad_1\hspace{-0.8pt}\ldots\gammad_\jb}(X,\Theta, \Thetab)\,.}[GeneralHomogeneity]

Let us pause for a moment to appreciate the importance of the result
\ThreePFAgen. The arbitrariness of the three-point function is now entirely
contained in the tensor $t$, which is only a function of the coordinates
$X,\Theta,\Thetab$, while the prefactor takes care of reproducing the
correct covariance properties at the first two points. Moreover, since
$\Theta,\Thetab$ are two component Grassmann spinors with R-charge $-1$,
$+1$, respectively, while $X$ and $x_{\bar{\imath}j}$ are neutral, it
follows that
\eqn{R = 0,\pm 1,\pm 2\,,}[]
with $R$ the R-charge of $\CO$. As we have already mentioned, and will
prove in Sec.~\ref{seccons}, the shortening condition \SupercurCons only
allows neutral supermultiplets, $R = 0$, to have a non-vanishing
correlation function.  In that case, the tensor $t$ is only allowed to
depend on the Grassmann variables through the combination
\eqn{
t^{\alphad\alpha;
  \lsp\betad\beta}{\!}_{\gamma_1\hspace{-0.8pt}\ldots\gamma_j;
  \lsp\gammad_1\hspace{-0.8pt}\ldots\gammad_\jb}( X, \Theta, \Thetab) = t^{\alphad\alpha;
  \lsp\betad\beta}{\!}_{\gamma_1\hspace{-0.8pt}\ldots\gamma_j;
  \lsp\gammad_1\hspace{-0.8pt}\ldots\gammad_\jb}( X,\Xb)\,,\qquad
  \Xb^\mu=X^\mu+2i\llsp\Theta\sigma^\mu\Thetab\,.
\lsp}[Xbar]
Moreover, from the invariance of the three-point function under
$z_1\leftrightarrow z_2$ and $\alpha\alphad\leftrightarrow\beta\betad$,
still for $R = 0$, we get
\eqn{ t^{\alphad\alpha;
  \lsp\betad\beta}
  {\!}_{\gamma_1\hspace{-0.8pt}\ldots\gamma_j;\lsp\gammad_1
  \hspace{-0.8pt}\ldots\gammad_\jb}(X,\Xb)
  = t^{\betad\beta;\lsp\alphad\alpha}
  {\!}_{\gamma_1\hspace{-0.8pt}\ldots\gamma_j;\lsp\gammad_1
  \hspace{-0.8pt}\ldots\gammad_\jb}(-\Xb,-X)\,.}[symmJJ]
Finally, in the case of real operators, for instance spin-$\ell$
supermultiplets with zero R-charge, we have the condition
\eqn{t^{\alphad\alpha;
  \lsp\betad\beta}
  {\!}_{\gamma_1\hspace{-0.8pt}\ldots\gamma_\ell;\lsp\gammad_1
  \hspace{-0.8pt}\ldots\gammad_\ell}(X,\Xb)^\ast=
 t^{\alphad\alpha;
  \lsp\betad\beta}
  {\!}_{\gamma_1\hspace{-0.8pt}\ldots\gamma_\ell;\lsp\gammad_1
  \hspace{-0.8pt}\ldots\gammad_\ell}(\Xb,X)\,.}[reality]

\subsec{Index-free notation}
\label{IndexFree}

For practical computations it is very convenient to contract all free indices with auxiliary commuting spinors $\eta_{i},\,\etab_{i}$. In this notation the three-point function reads
\eqn{\vev{\CJ(\etap_1,\etapb_1,z_1)\lsp\CJ(\etap_2,\etapb_2,z_2)\lsp
\CO_{j,\jb}(\eta_3,\etab_3,z_3)}=
\frac{
  \etap_1\xup_{1\bar{3}}\partial_{\etab_1}\lsp
  \partial_{\eta_1}\xup_{3\bar{1}}\etapb_1\lsp
  \etap_2\xup_{2\bar{3}}\partial_{\etab_2}\lsp
  \partial_{\eta_2}\xup_{3\bar{2}}\etapb_2
}{
  (x_{\bar{3}1}{\lnsp}^2\lsp x_{\bar{1}3}{\lnsp}^2\lsp
  x_{\bar{3}2}{\lnsp}^2\lsp x_{\bar{2}3}{\lnsp}^2)^2
}\lsp
t(\eta_{i},\etab_{i},X,\Theta,\Thetab)\,,
}[ThreePFEtagen]
where
\eqn{\begin{gathered}
\CJ(\eta,\etab,z)=\eta^\alpha\lsp\etab^{\llsp\alphad}
\CJ_{\alpha\alphad}(z)\,,\qquad
\CO_{j,\jb}(\eta,\etab,z)=\eta^{\alpha_1}\cdots\eta^{\alpha_j}\lsp
\etab^{\llsp\alphad_1}\cdots\etab^{\llsp\alphad_\jb}
\COindb{\alpha}(z)\,,\\
t(\eta_{i},\etab_{i},X,\Theta,\Thetab)=
\eta_{1\llsp\alpha}\lsp\etab_{1\llsp\alphad}\lsp\eta_{2\llsp\beta}\lsp
\etab_{2\llsp\betad}\lsp\lsp\eta_{3}^{\gamma_1}\cdots\eta_3^{\gamma_j}
\etab_{3}^{\gammad_1}\cdots\etab_3^{\gammad_\jb}
\lsp t^{\alphad\alpha;
\lsp\betad\beta}{\!}_{\gamma_1\hspace{-0.8pt}\ldots\gamma_j;
\lsp\gammad_1\hspace{-0.8pt}\ldots\gammad_\jb}(X,\Theta,\Thetab)\,.
\end{gathered}}[EtaNotation]
and derivatives with respect to $\eta$ obey
$\partial_{\eta^\alpha}\eta^\beta=\delta_{\alpha}{\!}^\beta$,
$\partial_{\eta^\alpha}=-\epsilon_{\alpha\beta}\lsp\partial_{\eta_\beta}$.

The problem is now reduced to finding the most general form of
$t(\eta_i,\etab_i,X,\Theta,\Thetab)$. This can be achieved using the
building blocks\footnote{This definition for the notation with square
brackets differs from the one used in Appendix~\ref{appendix:structureEven} by a replacement $X\to U$, defined later in \eqref{Udef}. The choice made here is less convenient for writing down the structures explicitly.}
\eqn{
\begin{gathered}
  \Uc{i}{\jmath} = \frac{\eta_i\lnsp\text{X}\etab_j}{|X|} \,,\quad \Uc{\Theta}{\Theta} = \frac{\Theta\text{X}\Thetab}{X^2}\,,  \quad   \Ec{i}{j} = \eta_i \eta_j\,, \quad \Ecb{\imath}{\jmath} = \bar{\eta}_i \bar{\eta}_j\,,\\
  \Ec{\Theta}{j} = \frac{\Theta \eta_j}{|X|^{1/2}} \,,\quad
    \Ecb{\Theta}{\jmath} = \frac{\Thetab \etab_j}{|X|^{1/2}} \,,\quad
    \Uc{j}{\Theta} = \frac{\eta_i\lnsp\text{X}\Thetab}{|X|^{3/2}}\,, \quad \Uc{\Theta}{\jmath} = \frac{\Theta\lnsp\text{X}\etab_j}{|X|^{3/2}}\,.
\end{gathered}}[Shortcuts]
In addition, the homogeneity property \GeneralHomogeneity can now we
written as
\eqn{
t(\eta_{1,2},\kappa\lsp\eta_3,\etab_{1,2},\bar{\kappa}\lsp\etab_3,\lambda\bar{\lambda}\lsp
X,\lambda\Theta,\bar{\lambda}\Thetab) =
\lambda^{\frac23(2q+\bar{q}-9)}\bar{\lambda}^{\frac23(q+2\bar{q}-9)}
\kappa^j\bar{\kappa}^\jb\, t(\eta_{i},\etab_{i},X,\Theta,\Thetab)\,,
}[homogeneityEta]
while the symmetry property \symmJJ for the first two points reads
\eqn{
t(\eta_{1,2,3},\etab_{1,2,3},X,\Xb)=
t(\eta_{2,1,3},\etab_{2,1,3},-\Xb,-X)
\,.
}[symm12Eta]
Lastly, recalling that complex conjugation swaps the order of fermions, the
reality condition \reality becomes
\eqn{
t(\eta_{i},\etab_{i},X,\Xb)^* = t(\etab_{i},\eta_{i},\Xb,X)\,.
}[realityEta]

Even though we have drastically simplified the problem, finding a complete
basis of tensor structures is still a non-trivial task. In order to
circumvent the issue of dealing with spinor identities but, at the same
time, be sure we do not miss any contributions, it is important to derive
separately the expected number of independent tensor structures appearing
in \ThreePFEtagen. This counting is performed in the following section.
After that we construct a complete basis for the cases of interest---namely
three-point functions of operators with vanishing R-charge---by providing
an equal number of independent tensor structures. Their independence can be
easily proven by setting to zero the Grassmann coordinates $\theta_{1,2}$
and $\thetab_{1,2}$ and matching with the non-supersymmetric three-point
functions reviewed in Sec.~\ref{Warmup}.  The tensor structures
associated to three-point functions of operators with non-zero R-charge can
be read from the \emph{Mathematica} notebook attached to this submission.
They are constructed in the same way, and their number agrees with the
counting of the next section as well. We also checked their linear
independence by replacing numerical values for the various quantities that
appear.

\subsec{Counting supersymmetric tensor structures}
\label{Counting}

In this section we obtain a group theoretical counting of the independent
number of tensor structures appearing in \ThreePFEtagen along the same
lines as \cite{Kravchuk:2016qvl}.

Let us analyze first the case of an operator $\CO_{j,\jb}$ with zero
R-charge. We can start by dividing the function
$t(\eta_i,\etab_i,X,\Theta,\Thetab)$ into three parts. The first part
contains neither $\Theta$ nor $\Thetab$; it is thus built with
$\Ec{i}{j},\,\Uc{i}{\jmath},\,\Ecb{\imath}{\jmath}$ only. The second part
is analogous to the previous one but with an overall $\Theta^2\Thetab^2$
factor. The third part is instead built with exactly one $\Theta$ and one
$\Thetab$.
In order to enumerate the structures in the first part we can simply follow
a standard approach for non-supersymmetric CFTs.
One possible way is to choose a conformal
frame~\cite{Mack:1976pa,Osborn:1993cr,Kravchuk:2016qvl} that fixes all
bosonic coordinates and breaks $\text{Spin}(2,4) \to
\text{Sp}(2,\mathbb{R})$. After restricting the polarizations $\eta_i$ and
$\etab_i$ to this subgroup, $\eta_{i\alpha}$ and $\tilde{\eta}_{i\alpha}
\equiv \text{X}_{\alpha\alphad}\etab_i^\alphad$ transform in the same
representation. Therefore we are allowed to make the
contractions\footnote{The contractions are made with
$\epsilon_{\alpha\beta}$ and $\epsilon^{\alpha\beta}$. All indices are
undotted at this point.}
\eqn{
\eta_i\eta_j\,,\qquad\eta_i\tilde{\eta}_j\,,\qquad\tilde{\eta}_i\tilde{\eta}_j\,,
}[]
which can be easily lifted in a one-to-one way to
$\Ec{i}{j},\,\Uc{i}{\jmath}$ and $\Ecb{\imath}{\jmath}$. For the purpose of
the subsequent arguments it is better to state this reasoning backwards.
The problem we need to address is the counting of independent structures
built out of products of $\Ec{i}{j},\,\Uc{i}{\jmath}$ and
$\Ecb{\imath}{\jmath}$, modded by all identities stemming from
$\epsilon_{\alpha[\beta}\epsilon_{\gamma\delta]}=0$. This is mathematically
equivalent to counting $\text{Sp}(2,\mathbb{R})$ invariant tensors built
out of $\eta_i$ and $\tilde{\eta}_i$, which are in one-to-one
correspondence with the tensor structures in $
\vev{\CJ_{1,1}(\eta_1,\etab_1,x_1)\CJ_{1,1}(\eta_2,\etab_2,x_2)
\CO_{j,\bar{\jmath}}(\eta_3,\etab_3,x_3)}$.

The second part presents no difference apart from the trivial
$\Theta^2\Thetab^2$ overall factor. The third part, instead, can be
interpreted in the following way: since there is only one $\Theta$ and only
one $\Thetab$ we can ignore the fact that they anticommute and replace them
by a fourth pair of polarizations $\eta_4,\,\etab_4$. In the same way as we
argued before, the enumeration of all structures is equivalent to the
enumeration of $\text{Sp}(2,\mathbb{R})$ invariants built out of $\eta_i$
and $\tilde{\eta}_i$, where now $i=1,\ldots,4$. The claim is that these are
in one-to-one correspondence with a three-point function where the third
operator transforms under the reducible representation of
$\text{SL}(2,\mathbb{C})$ given by
$(\tfrac{1}{2}j,\tfrac{1}{2}\jb)\otimes(\tfrac{1}{2},\tfrac{1}{2})$.\footnote{The
choice of attaching the polarizations $\eta_4,\,\etab_4$ to the third
operator is arbitrary and does not affect the result. In this case it is
convenient because we want to keep manifest the permutation symmetry in the
first two points.} This follows because irreducible representations are
tensors of the form
$\CO_{(\alpha_1\ldots\alpha_j);\llsp(\alphad_1\ldots\alphad_{\jb})}(x)$,
where parentheses denote symmetrization. In the index-free notation this
condition is enforced by contracting all indices with the same $\eta$.
Tensors not corresponding to irreducible representations must be contracted
with independent polarizations. For example, the following operator belongs
to $(\tfrac{1}{2}j,\tfrac{1}{2}\jb)\otimes(\tfrac{1}{2},\tfrac{1}{2})$,
\eqna{
\begin{split}
\CO_{(\alpha_1\ldots\alpha_j)\beta;\llsp(\alphad_1\ldots\alphad_{\jb})\betad}(x) \quad \to\quad
&\CO(x,\eta_1,\eta_2,\etab_1,\etab_2) \\
&\qquad\equiv
\eta_1^{\alpha_1}\cdots\eta_1^{\alpha_j}
\etab_1^{\alphad_1}\cdots\etab_1^{\alphad_\jb}
\eta_2^{\beta}\etab_2^{\betad}\lsp
\CO_{(\alpha_1\ldots\alpha_j)\beta;\llsp(\alphad_1\ldots\alphad_{\jb})\betad}(x)\,.
\end{split}
}

Now we are ready to perform the actual counting. In order to do so we will use the main formula derived in \cite{Kravchuk:2016qvl}
\eqn{
  N=\left(\mathrm{Res}^{\text{SO}(1,3)}_{\text{SO}(1,2)}
  \,\rho_1\otimes\rho_2\otimes\rho_3\right)^{\text{SO}(1,2)}\,.
}[grouptheo]
Here $\rho_1 = \rho_2 = (\tfrac{1}{2},\tfrac{1}{2})$ and $\rho_3$ is the spin representation of the third operator. The notation $\mathrm{Res}^G_H$ indicates the restriction of a representation of $G$ to a representation of $H\subseteq G$, the superscript $(\rho)^H$ denotes the $H$-singlets in $\rho$. We assume that this formula generalizes for $\rho_3$ not irreducible.
Moreover, as remarked in \cite{Kravchuk:2016qvl}, $\mathrm{Res}^G_H$
commutes with the tensor product. A last ingredient is necessary, namely
the permutation symmetry of the first two points. This is taken care of in
\cite{Kravchuk:2016qvl} as well. It is sufficient to replace
$\rho_1\otimes\rho_2$ by $\mathrm{S}^2\lsp \rho_1$ if $j$ is even and by
$\wedge^2 \lsp\rho_1$ if $j$ is odd, where $\mathrm{S}^2$ and $\wedge^2$ denote respectively the symmetrized square and the exterior square of representations.

Assuming, now, $j-\jb$ even,\footnote{If $j-\jb$ is odd the result is trivially zero.} we can write down the formulae
\eqna{
N(j,\jb)^{(j\text{ even})} &= 2\lsp \mathrm{Res}\left(\mathrm{S}^2\lsp(\tfrac{1}{2},\tfrac{1}{2}) \otimes(\tfrac{1}{2}j,\tfrac{1}{2}\jb)\right) + \mathrm{Res}\left(\wedge^2\lsp(\tfrac{1}{2},\tfrac{1}{2})\otimes (\tfrac{1}{2}j,\tfrac{1}{2}\jb)\otimes (\tfrac{1}{2},\tfrac{1}{2})\right) \,,\\
N(j,\jb)^{(j\text{ odd})} &= 2 \lsp \mathrm{Res}\left(\wedge^2\lsp(\tfrac{1}{2},\tfrac{1}{2})\otimes (\tfrac{1}{2}j,\tfrac{1}{2}\jb)\right) +\mathrm{Res}\left(\mathrm{S}^2\lsp(\tfrac{1}{2},\tfrac{1}{2})\otimes (\tfrac{1}{2}j,\tfrac{1}{2}\jb)\otimes (\tfrac{1}{2},\tfrac{1}{2})\right)\,,
}[nStructures]
where we abbreviated $\mathrm{Res}^{\text{SO}(1,3)}_{\text{SO}(1,2)}$ with $\mathrm{Res}$
and a superscript $\text{SO}(1,2)$ in all terms is understood. We denote the
number of independent structures in the three-point function
$\vev{\CJ_{1,1}\CJ_{1,1}\CO_{j,\jb}}$ with $N(j,\jb)$. The factor of ``2''
counts the first and second part. The second term comes from putting
$\rho_3 =
(\tfrac{1}{2},\tfrac{1}{2})\otimes(\tfrac{1}{2}j,\tfrac{1}{2}\jb)$ and
corresponds to the third part. Notice that when the first term has the
$\mathrm{S}^2$ product the second has the $\wedge^2$ product and vice
versa. This is because the tensor product $(\tfrac{1}{2}j,\tfrac{1}{2}\jb)\otimes(\tfrac{1}{2},\tfrac{1}{2})$ contains representations with different parity w.r.t. $(\tfrac{1}{2}j,\tfrac{1}{2}\jb)$. The result can be computed with the well known relations
\eqna{
\mathrm{Res}(\tfrac{1}{2}j,\tfrac{1}{2}\jb) = \bigoplus_{\ell = \frac{1}{2}|j-\jb|}^{\frac{1}{2}(j+\jb)} \underline{\ell}\;,\qquad
\underline{\ell} \otimes \underline{\ell}' = \bigoplus_{k =
|\ell-\ell'|}^{\ell+\ell'} \underline{k}\;,
}[tensorandres]
where $\underline{\ell}$ indicates the spin-$\ell$ representation of
$\text{SO}(1,2)$.
Finally the (anti)symmetrized products are given by
\eqna{
&\mathrm{S}^2\underline{\ell} = \bigoplus_{k =
2\ell \text{ mod } 2}^{2 \ell} \underline{k}\;,\qquad \wedge^2\underline{\ell} = \bigoplus_{k =
2\ell+1 \text{ mod } 2}^{2 \ell} \underline{k}\;,\\
&\begin{aligned}
\mathrm{S}^2 (\tfrac{1}{2}j,\tfrac{1}{2}\jb) &= (\mathrm{S}^2 \tfrac12
j,\mathrm{S}^2 \tfrac12 \jb)\oplus (\wedge^2 \tfrac12 j,\wedge^2 \tfrac12
\jb)\,, \\
\wedge^2 (\tfrac{1}{2}j,\tfrac{1}{2}\jb) &= (\mathrm{S}^2 \tfrac12
j,\wedge^2 \tfrac12 \jb)\oplus (\wedge^2 \tfrac12 j,\mathrm{S}^2 \tfrac12
\jb)\,,
\end{aligned}}[]
where $\text{S}^2$ and $\wedge^2$ inside the parenthesis $(\frac12 j,\frac12 \jb)$, stand for the direct sum of all possible pairs of the resulting irreps.

Collecting all the above results, the numbers of independent tensors structures consistent with permutation symmetry read
\eqn{
\begin{aligned}
N(\ell,\ell)^{(\ell\text{ even})} &= 16\,,\qquad &
N(\ell,\ell)^{(\ell\text{ odd})} &= 16\,,\\
N(\ell+2,\ell)^{(\ell\text{ even})} &= 10\,,\qquad &
N(\ell+2,\ell)^{(\ell\text{ odd})} &= 13\,,\\
N(\ell+4,\ell)^{(\ell\text{ even})} &= 4\,,\qquad &
N(\ell+4,\ell)^{(\ell\text{ odd})} &= 4\,.
\end{aligned}
}[ResultStructures]

Let us now consider the case of a supermultiplet $\CO_{j,\jb}$ with
non-zero R-charge. In this case  the superconformal primary does not
contribute to the three-point function. As a consequence the structures in
$t(\eta_i,\etab_i,X,\Theta,\Thetab)$ contain  an overall $\Theta$,
$\Thetab$, $\Theta^2$ or $\Thetab^2$. If the R-charge is $\pm 2$ the
problem is readily solved by multiplying the non-supersymmetric three-point
functions by $\Theta^2$ or $\Thetab^2$. The counting is therefore the same
as in Sec.~\ref{Warmup}.

When instead the
R-charge is $\pm 1$ (say $1$ for simplicity) we can derive a similar
formula as in Eq.~\nStructures. Here the structures can be divided into two parts, the
first proportional to $\Theta$ and the second proportional to
$\Theta^2\Thetab$. In both cases the free fermionic variable can be
interpreted as an extra $\eta_4$ or $\etab_4$ contracting a reducible
operator $\CO$ belonging to, respectively,
$(\tfrac{1}{2}j,\tfrac{1}{2}\jb)\otimes(\tfrac{1}{2},0)$ or
$(\tfrac{1}{2}j,\tfrac{1}{2}\jb)\otimes(0,\tfrac{1}{2})$. Therefore,
taking $j-\jb$ odd\footnote{The even case can be treated similarly and it trivially gives zero.} one finds
\eqna{
N(j,\jb)^{(j\text{ even})} &= \mathrm{Res}\left(\mathrm{S}^2\lsp(\tfrac{1}{2},\tfrac{1}{2}) \otimes (\tfrac{1}{2}j,\tfrac{1}{2}\jb)\otimes (0,\tfrac{1}{2})\right)+ \mathrm{Res}\left(\wedge^2\lsp(\tfrac{1}{2},\tfrac{1}{2}) \otimes (\tfrac{1}{2}j,\tfrac{1}{2}\jb)\otimes  (\tfrac{1}{2},0)\right) \,,\\
N(j,\jb)^{(j\text{ odd})} &= \mathrm{Res}\left(\wedge^2\lsp(\tfrac{1}{2},\tfrac{1}{2})\otimes (\tfrac{1}{2}j,\tfrac{1}{2}\jb)\otimes (0,\tfrac{1}{2})\right)+ \mathrm{Res}\left(\mathrm{S}^2\lsp(\tfrac{1}{2},\tfrac{1}{2}) \otimes (\tfrac{1}{2}j,\tfrac{1}{2}\jb)\otimes  (\tfrac{1}{2},0)\right) \,.
}[nStructuresOther]
Again notice that the products $\mathrm{S}^2$ and $\wedge^2$ are inverted in the
two terms. The counting now gives
\eqn{
\begin{aligned}
N(\ell\pm 1,\ell)^{(\ell\text{ even})} &= 10\,,\qquad &
N(\ell\pm 1,\ell)^{(\ell\text{ odd})} &= 10\,,\\
N(\ell\pm 3,\ell)^{(\ell\text{ even})} &= 5\,,\qquad &
N(\ell\pm 3,\ell)^{(\ell\text{ odd})} &= 5\,,\\
N(\ell\pm 5,\ell)^{(\ell\text{ even})} &= 1\,,\qquad &
N(\ell\pm 5,\ell)^{(\ell\text{ odd})} &= 1\,.
\end{aligned}
}[]

From this analysis we can deduce a recipe for counting structures of more
general superconformal three-point functions. Let us assume that all
operators are different and belong to $\text{SO}(1,3)$ representations $\rho_1$, $\rho_2$ and $\rho_3$ respectively. The cases with non-trivial permutation symmetries
can be treated similarly as above. Every function
$t(\eta_i,\etab_i,X,\Theta,\Thetab)$ can contain a subset of the monomials
\eqn{
\Theta^0\Thetab^0\;,\quad\;\Theta^\alpha\;,\quad\; \Thetab^2 \Theta^\alpha\;,\quad\;
\Thetab^\alphad\;,\quad\; \Theta^\alpha\Thetab^\alphad\;,\quad\; \Theta^2\Thetab^\alphad\;,\quad\; \Theta^2\;,\quad\; \Thetab^2\;,\quad\; \Theta^2 \Thetab^2\,.
}[Ordersintheta]
Which ones are present depends on the R-charges of the operators
$\CO_1,\,\CO_2,\,\CO_3$, which we will call $r_1,\,r_2,\,r_3$. Define
\eqn{
\delta = r_3 - r_1 - r_2\,.
}[]
The possible values are $\delta= \pm 2,\,\pm 1,\,0$. Let us denote as $N_X(\rho_1,\rho_2,\rho_3)$ the number of structures of a given order $X$ in $\Theta,\Thetab$, where $X$ is any monomial in \Ordersintheta. Following the analysis above we have
\eqna{
N_1(\rho_i) = N_{\Theta^2\Thetab^2}(\rho_i)  = N_{\Theta^2}(\rho_i)  = N_{\Thetab^2}(\rho_i)  &= \mathrm{Res} \,\rho_1\otimes\rho_2\otimes\rho_3\;,\\
N_{\Theta}(\rho_i) = N_{\Theta\Thetab^2}(\rho_i) &= \mathrm{Res} \,\rho_1\otimes\rho_2\otimes\rho_3\otimes(\tfrac{1}{2},0)\;,\\
N_{\Thetab}(\rho_i) = N_{\Theta^2\Thetab}(\rho_i) &= \mathrm{Res} \,\rho_1\otimes\rho_2\otimes\rho_3\otimes(0,\tfrac{1}{2})\;,\\
N_{\Theta\Thetab}(\rho_i) &= \mathrm{Res} \,\rho_1\otimes\rho_2\otimes\rho_3\otimes(\tfrac{1}{2},\tfrac{1}{2})\;.\\
}[]
where again $\mathrm{Res} \equiv
\mathrm{Res}^{\text{SO}(1,3)}_{\text{SO}(1,2)}$ and a superscript
$\text{SO}(1,2)$ in all terms is understood. Then the general formula for the number $N(\rho_1,\rho_2,\rho_3;\delta)$ of tensor structures in the three-point function $\langle\CO_1\CO_2\CO_2\rangle$ may be written as
\eqn{
N(\rho_1,\rho_2,\rho_3;\delta) = \left\lbrace
\begin{aligned}
&2N_1(\rho_1,\rho_2,\rho_3) + N_{\Theta\Thetab}(\rho_1,\rho_2,\rho_3) & \delta &= 0\,,\\
&N_1(\rho_1,\rho_2,\rho_3) & \delta &= \pm 2\,,\\
& N_\Theta(\rho_1,\rho_2,\rho_3) +N_\Thetab(\rho_1,\rho_2,\rho_3)  & \delta &= \pm 1\,.\\
\end{aligned}\right.
}[grouptheogen]

\subsec{Conserved tensor structures}[seccons]
\par To conclude this section we will address the issue of conservation. In particular we will study the consequence of \SupercurCons on a general three-point function $\langle \CJ\CJ\CO\rangle$. For simplicity we will omit the superspace coordinate dependence. The constraints we need to impose are
\twoseqn{
\langle D^\alpha \CJ_{\alpha\alphad}\lsp \CJ_{\beta\betad}\lsp \CO_{j,\jb}\rangle = 0\,,
}[DJJO]{
\langle \Db^\alphad \CJ_{\alpha\alphad}\lsp \CJ_{\beta\betad}\lsp \CO_{j,\jb}\rangle = 0\,.
}[DbJJO][DDbJJO]
These conditions are not independent; in fact there are linear relations
between them. First we can observe that taking the derivative $\Db$ at the
second point of \DJJO and the derivative $D$ at the second point of \DbJJO
give the same result, modulo permuting the first two operators,
\eqn{
\langle D^\alpha \CJ_{\alpha\alphad}\lsp \Db^\betad\CJ_{\beta\betad}\lsp \CO_{j,\jb}\rangle = \langle \Db^\alphad \CJ_{\alpha\alphad}\lsp D^\beta\CJ_{\beta\betad}\lsp \CO_{j,\jb}\rangle \big|_{1\leftrightarrow 2 }\,.
}[]
Moreover, by taking $D$ of \DJJO and permuting points $z_1$ and $z_2$ we
obtain identically zero. The same holds if we take $\Db$ of \DbJJO. The prescription to count the number of conserved tensor structures~\cite{Kravchuk:2016qvl} is to take the number of non-conserved tensor structures, subtract all degrees of freedom contained in the equations \DDbJJO and add back all linear relations between such equations. The complication with supersymmetry is that a superspace equation decomposes into a certain number of ordinary bosonic equations by projecting on the various terms in \Ordersintheta. This depends on the R-charge of $\CO$. Let us start assuming that $
\CO$ is real. The conservation conditions impose a number of constraints equal to the number of tensor structures present in  \DDbJJO. This number is given by\footnote{We denote the various representations in $N_X(\ldots)$ in the following way:
\eqn{
  \CJ = (\tfrac12,\tfrac12)\,,\quad D\CJ = (0,\tfrac12)\,,\quad
\Db\CJ = (\tfrac12,0)\,,\quad\CO = (\tfrac{1}{2}j,\tfrac{1}{2}\jb)\,.}[]
}
\eqn{
N_\Thetab(D\CJ,\CJ,\CO) + N_{\Theta\Thetab^2}(D\CJ,\CJ,\CO)+
N_\Theta(\Db\CJ,\CJ,\CO) + N_{\Theta^2\Thetab}(\Db\CJ,\CJ,\CO)\,.
}[CountingCons]
Even though $N_\Thetab = N_{\Theta^2\Thetab}$, etc., we keep them distinct to track down the various
contributions.
As anticipated, however, not all the tensor structures in \DDbJJO give a non trivial constraint. This is a consequence of the fact that the three-point functions $D \DJJO$ and $\Db \DbJJO$ are made of identical operators. To keep this into account one must subtract from \CountingCons the numbers
\twoseqn{
N_{\Thetab^2}(D\CJ,D\CJ,\CO)\,,
}[DJDJO]
{
N_{\Theta^2}(\Db\CJ,\Db\CJ,\CO)\,.
}[DbJDbJO][DDbJDDbJO]
Similarly, given the relation $\Db \DJJO \sim D \DbJJO$, we should na\^ively subtract from \CountingCons  the number
\eqn{
N_1(D\CJ,\Db\CJ,\CO) + N_{\Theta\Thetab}(D\CJ,\Db\CJ,\CO) + N_{\Theta^2\Thetab^2}(D\CJ,\Db\CJ,\CO)\,.
}[DJDbJO]
However, the above expression would give rise to an over-counting: the
conditions given by $N_{\Theta^2\Thetab^2}(D\CJ,\Db\CJ,\CO)$ and by
$N_1(D\CJ,\Db\CJ,\CO)$ are dependent.  Indeed, by using a suitable
representation of the differential operators,\footnote{See
Appendix~\ref{Conventions}, specifically Eq.~\eqref{OsbornDefCDCDb}.} one
can show that the terms $\Theta^2\Thetab^2$ cannot be generated by applying
$D$ and $\Db$ on $\langle \CJ\CJ\CO\rangle$.  The correct counting is then
\eqna{
N(\CJ,\CJ,\CO;0)^{(\text{cons.})}&=N(\CJ,\CJ,\CO;0) - N(D\CJ,\CJ,\CO;1) - N(\Db\CJ,\CJ,\CO;-1) \\&\hspace{.44cm}
+ N(D\CJ,D\CJ,\CO;2) + N(\Db\CJ,\Db\CJ,\CO;-2)\\&\hspace{.44cm} + N_1(D\CJ,\Db\CJ,\CO) + N_{\Theta\Thetab}(D\CJ,\Db\CJ,\CO)\,.
}[]
In addition, since the currents $\mathcal J$ are identical, we need to take into account the permutation symmetry as we explained in the previous section by replacing the
product $\rho_1\otimes\rho_2$ by either $\mathrm{S}^2\rho_1$ or
$\wedge^2\rho_1$. There is a subtlety in the (anti)symmetrization of two
$D\CJ$'s or two $\Db\CJ$'s: these operators get an extra minus due to their fermionic nature. Thus for $\ell$ even (odd) we must take the $\mathrm{S}^2$ ($\wedge^2$) product in $N(D\CJ,D\CJ,\CO;-2)$ and $N(\Db\CJ,\Db\CJ,\CO;2)$.
Explicitly this formula yields
\eqn{
\begin{aligned}
N(\ell,\ell)^{(\ell\text{ even, cons.})} &= 2\,,\qquad &
N(\ell,\ell)^{(\ell\text{ odd, cons.})} &= 2\,,\\
N(\ell+2,\ell)^{(\ell\text{ even, cons.})} &= 1\,,\qquad &
N(\ell+2,\ell)^{(\ell\text{ odd, cons.})} &= 2\,,\\
N(\ell+4,\ell)^{(\ell\text{ even, cons.})} &= 1\,,\qquad &
N(\ell+4,\ell)^{(\ell\text{ odd, cons.})} &= 1\,,
\end{aligned}
}[ResultStructuresCons]
where, as before, $N(j,\jb)$ is a shorthand for $N\big(\CJ,\CJ,(\frac12 j,\frac12 \jb);0\big)$.

We can similarly obtain the respective formulas when $\CO$ has non-zero R-charge. Without loss of generality we take the R-charge to be negative.\footnote{The other case can be obtained by complex conjugation. In order to prove the same formula for $R=1$ we would need a representation of the differential operators where $\Db_\alphad \to \partial/\partial \Theta^\alpha$, namely \eqref{OsbornDefCDCDb2}.} Skipping the details of the derivation we show the answer for $R=-
1$,
\eqna{
N(\CJ,\CJ,\CO;-1)^{(\text{cons.})}&=N(\CJ,\CJ,\CO;1) - N(\Db\CJ,\CJ,\CO;-2) - N_1(D\CJ,\CJ,\CO) \\&\hspace{.44cm}- N_{\Theta\Thetab}(D\CJ,\CJ,\CO)
+N_{\Theta}(D\CJ,D\CJ,\CO) + N_{\Thetab}(D\CJ,\Db\CJ,\CO)\,,
}[RCone]
and for $R=- 2$,
\eqna{
N(\CJ,\CJ,\CO;-2)^{(\text{cons.})}&=N(\CJ,\CJ,\CO;-2) - N_{\Thetab}(D\CJ,\CJ,\CO) + N_1(D\CJ,D\CJ,\CO)\,.
}[RCtwo]
In all cases with non-zero R-charge \RCone and \RCtwo yield non-positive
results. Therefore we conclude that there are no structures allowed after
conservation, as mentioned in Sec.~\ref{General}.

\newsec{Ward identities and their solution}
\label{Ward}

In this section we explore the consequences of the Ferrara--Zumino
shortening conditions, Eq.~\SupercurCons. For each of the cases
\ref{caseA}--\ref{caseC} we write the correlator in superspace in terms of
tensor structures satisfying the conditions \GeneralHomogeneity,  \symmJJ
and eventually \reality.  One can straightforwardly check their
independence, and since they match in number the predictions of
Sec.~\ref{Counting}, they form a basis of superconformal three-point functions.

\subsec{Case A:~\texorpdfstring{$(\frac12\ell,\frac12\ell)$}
  {(l/2,l/2)} operators}
  \label{caseAsection}
We begin by  considering the case where $\CO$ is a spin-$\ell$ Lorentz
representation with zero R-charge. This means that $j=\jb=\ell$, i.e.\
$\CO$ has $\ell$ dotted and $\ell$ undotted indices, which are symmetrized
separately, and $q=\qb=\frac12\Delta$. The unitarity bound \GenUniBoundA
simply becomes
\eqn{\Delta\geq \ell+2\,,}[UniBoundA]
which agrees with the usual non-supersymmetric unitarity
bounds, Eq.~\NonSusyUnitarity.

\subsubsec{Even \texorpdfstring{$\ell$}{l}}
When all Grassmann variables are set to zero, then there are four possible
parity-even structures in $t_{\text{A}}$ for $\ell$
even~\cite{Echeverri:2015rwa}.  The Ward identity that follows from
\SupercurCons will relate these structures, leaving, in the end, at most
two independent parity-even structures~\cite{Echeverri:2015rwa}. In our
case there may also be structures that become identically zero when all
Grassmann variables are set to zero.

It is easier to perform the Ward-identity analysis by first introducing
auxiliary commuting spinors $\eta_{i},\etab_{i}$, $i=1,2,3$, as
in~\EtaNotation. We can use these spinors to write the three-point function
in the form\foot{In order to have a consistent treatment of all cases we
have chosen to express $t_{\text{A}}^{(\ell\text{ even})}$ as a function of
$U,\Theta\Thetab$ instead of $X,\Xb$.}
\eqna{\vev{\CJ(\etap_1,\etapb_1,z_1)\lsp\CJ(\etap_2,\etapb_2,z_2)\lsp
\CO_\ell(\eta_3,\etab_3,z_3)}=
\frac{
  \etap_1\xup_{1\bar{3}}\partial_{\etab_1}\lsp
  \partial_{\eta_1}\xup_{3\bar{1}}\etapb_1\lsp
  \etap_2\xup_{2\bar{3}}\partial_{\etab_2}\lsp
  \partial_{\eta_2}\xup_{3\bar{2}}\etapb_2
}{
  (x_{\bar{3}1}{\lnsp}^2\lsp x_{\bar{1}3}{\lnsp}^2\lsp
  x_{\bar{3}2}{\lnsp}^2\lsp x_{\bar{2}3}{\lnsp}^2)^2
}\lsp
t_{\text{A}}(\eta_{i},\etab_{i},U,\Theta\Thetab)\,,
}[ThreePFAEta]
with the definition
\eqn{
U = \tfrac{1}{2}(X+\Xb)\,.
}[Udef]
As anticipated in Sec.~\ref{Counting}, we can parametrize $t_A$ in terms of 16  coefficients,
\eqna{
t_{\text{A}}^{(\ell\text{ even})}
(\eta_{i},\etab_{i},U,\Theta\Thetab)=
 \frac{ ( \eta_3\lnsp\Uup\etab_3)^{\ell-2}}{ U^{4-\Delta+\ell}}\lsp
 \left(\sum_{i=1}^{4} \TAEven{i} \left(A_i +B_i \lsp \xi^2\right) +
 \sum_{i=1}^{8} \TAEven[\llsp\prime\llsp]{i} C_i\right) \,,\qquad
 \xi^2 = \lsp\frac{\Theta^2\Thetab^2}{U^2}\,,
}[tAEvenSpinEta]
where the tensor structures $\TAEven{i}$ and $\TAEven[\llsp\prime\llsp]{i}$
are given in Appendix~\ref{appendix:structureEven}. When all the variables $\theta_i$ and $\thetab_i$ are set to zero, only the structures $\TAEven{i}$ survive, while all the others vanish. Hence, the coefficients $A_i$ must be related to the coefficients $\lambda^{(j)}$ introduced in Sec.~\ref{Warmup} for the case of traceless symmetric tensors ($k=0$). It is straightforward to find
\eqn{
\lambda^{(1)}=-(\AevenA+\AevenC-\AevenH)\,,\qquad
\lambda^{(2)}=-\lambda^{(3)}=-2\llsp \AevenH\,,\qquad
\lambda^{(4)}=\AevenF\,,\qquad
\lambda^{(5)}=\AevenA-\AevenC-\AevenH\,.}[AEvenZero]

The Ward identities following from \SupercurCons are satisfied if
\eqn{
\partial_{\eta_1}\!\mathcal{D}\,
t_{\text{A}}^{(\ell\text{ even})}
(\eta_{i},\etab_{i},U,\Theta\Thetab)=0\,.}[WardtAEven]
Equation \WardtAEven decomposes into fourteen independent equations for the
sixteen \emph{a priori} independent coefficients $\AevenA,\ldots,\AevenP$.\foot{For this it is crucial to
use identities that stem from
$\epsilon_{\alpha[\beta}\epsilon_{\gamma\delta]}=0$ and the corresponding
identity with dotted indices. Examples include the identities in
\cite[Appendix E]{Cuomo:2017wme}. In this work we circumvent the need to
impose such idetities by substituting numerical values for the various
quantities that appear. This is equivalent to working in the superconformal
frame.} We may express all coefficients in terms of $\AevenA$ and $\AevenF$
to find
\small
\begin{align}
\label{WardtAEvenSol}
\AevenC&=-\frac{(\Delta-\ell -6) (\Delta-\ell -2) }{3(\Delta-2)^2 - \ell(\ell+2)}\AevenA+\frac{(\Delta-\ell -6) (\Delta-\ell -4) }{4 (3(\Delta-2)^2 - \ell(\ell+2))}\AevenF\,,\notag\\
\AevenH&=-\frac{2 (\Delta-3) \ell  }{3(\Delta-2)^2 - \ell(\ell+2)}\AevenA-\frac{(\Delta-\ell -4) (3 \Delta+\ell -6) }{4 (3(\Delta-2)^2 - \ell(\ell+2))}\AevenF\,,\notag\\
\AevenB&=-\tfrac{1}{4} (\Delta-\ell -6) (\Delta+\ell -4) \AevenA\,,\notag\\
\AevenG&=\frac{8 (\Delta-3) (\ell -1) \ell  }{3(\Delta-2)^2 -
\ell(\ell+2)}\AevenA-\frac{(\Delta-\ell -4) (\Delta+\ell -2) (3(\Delta-4)^2
-\ell(\ell+2)) }{4 (3(\Delta-2)^2 - \ell(\ell+2))}\AevenF\,,\notag\\
\AevenD&=\frac{(\Delta-\ell -8) (\Delta-\ell -6) (\Delta-\ell -2) (\Delta+\ell -2) }{4 (3(\Delta-2)^2 - \ell(\ell+2))}\AevenA-\frac{(\Delta-\ell -8) (\Delta-\ell -6) (\Delta-\ell -4) (\Delta+\ell -2) }{16 (3(\Delta-2)^2 - \ell(\ell+2))}\AevenF\,,\notag\\
\AevenI&=\frac{\ell  (\Delta-\ell -6) ((\Delta-5)(\Delta+\ell)+8) }{2 (3(\Delta-2)^2 - \ell(\ell+2))}\AevenA+\frac{(\Delta-\ell -6) (\Delta-\ell -4) (\Delta+\ell -2) (3 \Delta+\ell -12) }{16 (3(\Delta-2)^2 - \ell(\ell+2))}\AevenF\,,\notag\\
\AevenE&=\frac{(\Delta-\ell -6) (\Delta-\ell -2) (\Delta+\ell -2) }{3(\Delta-2)^2 - \ell(\ell+2)}\AevenA-\frac{(\Delta-\ell -6) (\Delta-\ell -4) (\Delta+\ell -2) }{4 (3(\Delta-2)^2 - \ell(\ell+2))}\AevenF\,,\notag\\
\AevenJ&=\frac{4 (\Delta-3) (\ell -1) \ell  }{3(\Delta-2)^2 - \ell(\ell+2)}\AevenA+\frac{3 (\Delta-3) (\Delta-\ell -4) (\Delta+\ell -2) }{2 (3(\Delta-2)^2 - \ell(\ell+2))}\AevenF\,,\notag\\
\AevenK&=\AevenN=\AevenO=\AevenP=0\,,\notag\\
\AevenL&=\frac{2 \ell  (\Delta^2-4 \Delta-\ell ^2-2 \ell +6) }{3(\Delta-2)^2 - \ell(\ell+2)}\AevenA-\frac{(\ell +3) (\Delta-\ell -4) (\Delta+\ell -2) }{2 (3(\Delta-2)^2 - \ell(\ell+2))}\AevenF\,,\notag\\
\AevenM&=-\frac{2 \ell(\ell -1) (\Delta-\ell -6) }{3(\Delta-2)^2 - \ell(\ell+2)}\AevenA-\frac{3 (\Delta-\ell -6) (\Delta-\ell -4) (\Delta+\ell -2) }{4 (3(\Delta-2)^2 - \ell(\ell+2))}\AevenF\,.
\end{align}\normalsize
\newcommand{\WardtAEvenSol}{\eqref{WardtAEvenSol}\xspace}

The above expressions are divergent if we set $\Delta = 2$ and $\ell=0$.
The special value $\ell=0$ is discussed separately in
Appendix~\ref{appendix:WISpinZeroOne}. In that case there is then only one
undetermined coefficient, consistently with the results
of~\cite{Osborn:1998qu}.

\subsubsec{Odd \texorpdfstring{$\ell$}{l}}
For odd spins we may write
\eqna{
t_{\text{A}}^{(\ell\text{ odd})}
(\eta_{i},\etab_{i},U,\Theta\Thetab)&=
 \frac{ ( \eta_3\lnsp\Uup\etab_3)^{\ell-3}}{ U^{3-\Delta+\ell}}\lsp \Bigg(\sum_{i=1}^2 \TAOdd{i}\left(A_i+B_i \lsp \xi^2\right)+\sum_{i=1}^3 \TAOdd[\llsp\prime\llsp]{i}\left(C_i\lsp \zeta+D_i \lsp \zeta'\right)
  +\sum_{i=1}^{6}\TAOdd[\llsp\prime\prime\llsp]{i}\lsp E_i\Bigg) \,,\\
    \xi^2& = \lsp\frac{\Theta^2\Thetab^2}{U^2}\,,\qquad \zeta = \frac{\Theta\eta_3\lsp\Thetab\etab_3}{|U|}\,,\qquad \zeta' = \frac{\Theta \Uup \Thetab}{|U|^{3/2}}\lsp\eta_3\lnsp \Uup\etab_3\,,
}[tAOddSpinEta]
where $A_i,\,B_i,\,C_i,\,D_i,\,E_i$ are real constants and the tensor
structures $\TAOdd{i}, \TAOdd[\llsp\prime\llsp]{i}$ and
$\TAOdd[\llsp\prime\prime\llsp]{i}$ are given in
Appendix~\ref{appendix:structureEven}.

In the lowest component of the three-point function \ThreePFAEta for
general odd spin $\ell$ there is a parity-odd and a parity-even structure,
with respective coefficients denoted by
$\lambda^{(-)}$ and $\lambda^{(2)}$ in Sec.~\ref{Warmup}.
In terms of $\AoddA$ and $\AoddC$ of \tAOddSpinEta  we find
%
\eqn{\lambda^{(-)}=2\lsp i\llsp \AoddA\,,\qquad\lambda^{(2)}=-2\lsp \AoddC\,.
}[coeffsOddZero]

Again, the shortening condition \SupercurCons implies
\eqn{
\partial_{\eta_1}\!\mathcal{D}\,
t_{\text{A}}^{(\ell\text{ odd})}
(\eta_{i},\etab_{i},U,\Theta\Thetab)=0\,.}[WardtAOdd]
For generic $\Delta,\ell$ this gives fourteen independent equations for the
sixteen unknowns $\AoddA,\ldots,\AoddP$, and thus there are two
undetermined coefficients, just as in the even-spin case. If we choose
$\AoddA$ and $\AoddM$ as independent we have\small
\begin{align}
\label{WardtAOddSol}
\AoddC&=\AoddD=\AoddO=\AoddP=0\,,\notag\\
\AoddB&=-\tfrac{1}{4}  (\Delta-\ell -6) (\Delta+\ell -4)\AoddA\,,\notag\\
\AoddI&=\frac{2 (\Delta-\ell -6) (\Delta-\ell -2) (2 \Delta+\ell -4)}{
5(\Delta-3)(\Delta-1)-3(\ell-1)(\ell+3)}\AoddA +\frac{ (\Delta-\ell -6) (\Delta-\ell -4) (2 \Delta+\ell -4)}{4 (\Delta-2) (5(\Delta-3)(\Delta-1)-3(\ell-1)(\ell+3))}\AoddM\,,\notag\\
\AoddE&=-2 (\Delta-4) \AoddA\,,\notag\\
\AoddN&=-\frac{8  (\ell -1) (\Delta-\ell -2) (\Delta+\ell )}{5(\Delta-3)(\Delta-1)-3(\ell-1)(\ell+3)}\AoddA-\frac{ (\Delta-\ell -4) (5 \Delta^2+2 \Delta \ell -22 \Delta-\ell ^2-8 \ell +24)}{2 (\Delta-2) (5(\Delta-3)(\Delta-1)-3(\ell-1)(\ell+3))}\AoddM\,,\notag\\
\AoddJ&=-\frac{ (\Delta-\ell -8) (\Delta-\ell -6) (\Delta-\ell -2)}{5(\Delta-3)(\Delta-1)-3(\ell-1)(\ell+3)}\AoddA-\frac{ (\Delta-\ell -8) (\Delta-\ell -6) (\Delta-\ell -4)}{8 (\Delta-2) (5(\Delta-3)(\Delta-1)-3(\ell-1)(\ell+3))}\AoddM\,,\notag\\
\AoddF&=(\Delta-\ell -6)\AoddA \,,\notag\\
\AoddG&=\frac{ (\Delta-\ell -4) (\Delta+\ell )}{4 (5(\Delta-3)(\Delta-1)-3(\ell-1)(\ell+3))}\AoddM\notag\\
&\hspace{.44cm}+\frac{2  (\Delta^3-9 \Delta^2-\Delta \ell ^2-2 \Delta \ell +24 \Delta+5 \ell ^2+10 \ell -24)}{5(\Delta-3)(\Delta-1)-3(\ell-1)(\ell+3)}\AoddA\,,\notag\\
\AoddH&=\frac{ (\Delta-\ell -6) (\Delta-\ell -4) (2 \Delta+\ell -4)}{4 (\Delta-2) (5(\Delta-3)(\Delta-1)-3(\ell-1)(\ell+3))}\AoddM\notag\\
&\hspace{.44cm}-\frac{ (\Delta-\ell -6) (\Delta^2+2 \Delta \ell -4 \Delta-\ell ^2-10 \ell +8)}{5(\Delta-3)(\Delta-1)-3(\ell-1)(\ell+3)}\AoddA\,,\notag\\
\AoddK&=-\frac{ (\Delta-\ell -6) (\Delta-\ell -2) (\Delta+\ell )}{5(\Delta-3)(\Delta-1)-3(\ell-1)(\ell+3)}\AoddA-\frac{ (\Delta-\ell -6) (\Delta-\ell -4) (\Delta+\ell )}{8 (\Delta-2) (5(\Delta-3)(\Delta-1)-3(\ell-1)(\ell+3))}\AoddM\,,\notag\\
\AoddL&=\frac{2  (\ell -1) (3 \Delta^2-16 \Delta-\ell ^2-2 \ell +24)}{5(\Delta-3)(\Delta-1)-3(\ell-1)(\ell+3)}\AoddA\notag\\
&\hspace{.44cm}-\frac{ (\Delta-\ell -4) (5 \Delta^2+2 \Delta \ell -22 \Delta-\ell ^2-8 \ell +24)}{4 (\Delta-2) (5(\Delta-3)(\Delta-1)-3(\ell-1)(\ell+3))}\AoddM\,.
\end{align}\normalsize
\newcommand{\WardtAOddSol}{\eqref{WardtAOddSol}\xspace}
\vspace{-\baselineskip}

\noindent The result $\AoddC=0$ is expected because for conserved currents there is
only one parity-odd structure in $t_{\text{A}}^{(\ell\text{ odd})}$ if all
Grassmann variables are set to zero~\cite{Echeverri:2015rwa}.  If we set
$\ell=1$ and $\Delta=3$, corresponding to the third operator in the
three-point function being the supercurrent, we find a singularity in the
above expression. This is just an artifact of the choice of variables and
is resolved in Appendix~\ref{appendix:WISpinZeroOne}. In the same Appendix we also show the relation between the coefficients defined here and the anomaly coefficients $a$ and $c$.

\subsec{Case B:~\texorpdfstring{$(\frac12(\ell+2),\frac12\ell)$}
{((l+2)/2,l/2)} operators}
Here we start with
\eqna{
\vev{\CJ(\eta_1,\etab_1,z_1)\lsp\CJ(\eta_2,\etab_2,z_2)\lsp
\CO_{\ell+2,\ell}(\eta_3,\etab_3,z_3)}&\\
&\hspace{-2cm}=
\frac{
  \eta_1\xup_{1\bar{3}}\partial_{\etapb_1}\lsp
  \partial_{\etap_1}\xup_{3\bar{1}}\etab_1\lsp
  \eta_2\xup_{2\bar{3}}\partial_{\etapb_2}\lsp
  \partial_{\etap_2}\xup_{3\bar{2}}\etab_2
}{
  (x_{\bar{3}1}{\lnsp}^2\lsp x_{\bar{1}3}{\lnsp}^2\lsp
  x_{\bar{3}2}{\lnsp}^2\lsp x_{\bar{2}3}{\lnsp}^2)^2
}\lsp
t_{\text{B}}(\etap_{1,2},\eta_3,\etapb_{1,2},\etab_3,U,\Theta\Thetab)\,,
}[ThreePFB]
where $\CO$ has zero R-charge. The unitarity bound is
\eqn{\Delta\geq\ell+4\,.}[UniBoundB]
Similarly to \homogeneityEta and \eqref{symm12Eta} we must impose the
homogeneity property
\eqn{
t_{\text{B}}
(\eta_{1,2},\kappa\lsp\eta_3,\etab_{1,2},\bar{\kappa}\lsp\etab_3,\lambda\bar{\lambda}\lsp
U,\lambda\llsp\Theta\lsp\llsp \bar{\lambda}\llsp\Thetab) =
(\lambda\bar{\lambda})^{\Delta-6}\kappa^{\ell+2}\bar{\kappa}^\ell\, t_{\text{B}}
(\eta_{i},\etab_{i},U,\Theta\Thetab)\;.
}[]
The symmetry properties for the first two points is identical to
\eqref{symm12Eta}. However, unlike in case \ref{caseA},
we do not have a reality condition as in \realityEta.

\subsubsec{Even \texorpdfstring{$\ell$}{l}}\label{caseBEven}
For general even $\ell$, we can parametrize the correlator as
\eqn{t_{\text{B}}^{(\ell\text{ even})}
(\eta_{i},\etab_{i},U,\Theta\Thetab)=
 \frac{ ( \eta_3\lnsp\Uup\etab_3)^{\ell-1}}{ U^{5-\Delta+\ell}}\lsp
 \left(\sum_{i=1}^{2} \TBEven{i} \left(A_i +B_i \lsp \xi^2\right) +
 \sum_{i=1}^{6} \TBEven[\llsp\prime\llsp]{i} C_i\right) \,,\qquad
    \xi^2 = \lsp\frac{\Theta^2\Thetab^2}{U^2}\,,}[tBEvenSpin]
where the structures $\TBEven{i}$ and $\TBEven[\llsp\prime\llsp]{i}$ are
defined in Appendix~\ref{appendix:structureEven}. At the lowest oder in $\theta_i$, $\thetab_i$
only the structures $\TBEven{i}$ survive and they must reproduce the tensor structures $\mathcal T^{(i)}$ introduced in Sec.~\ref{Warmup}.
The relations between $\BevenC$ and $\BevenA$ of \tBEvenSpin and the $\lambda^{(i)}$ in \JJOnonsusy is
\eqn{\lambda^{(1)}=-\lambda^{(2)}=\BevenC\,,\qquad
\lambda^{(3)}=\lambda^{(4)}=-2\lsp \BevenA\,.}[BEvenZero]

The Ward identity this time requires two independent conditions,
\eqn{
\partial_{\eta_1}\!\mathcal{D}\,t_{\text{B}}^{(\ell\text{ even})}(\eta_{i},\etab_{i},U,\Theta\Thetab)=0\,,\qquad
\partial_{\etab_1}\!\overbar{\mathcal{D}}\,t_{\text{B}}^{(\ell\text{ even})}(\eta_{i},\etab_{i},U,\Theta\Thetab)=0
\,,}[WardtBEven]
due to the lack of a reality condition on $t_\text{B}$. This leads, for
generic $\Delta$ and $\ell$, to nine linear constraints for the ten
unknowns $\BevenC,\ldots, \BevenJ$. Solving in terms of  $\BevenC$ we
find
\eqn{\begin{aligned}
\BevenA&=-\frac{\Delta-\ell -6}{4(\Delta-2)}\BevenC \,,\qquad &
\BevenD&=-\frac{(\Delta-4)(\Delta-\ell-6) (\Delta+\ell -2)}{4 (\Delta-
2)}\BevenC\,,\\
\BevenB&=\frac{(\Delta-\ell -8) (\Delta-\ell -6) (\Delta+\ell -2)}{16
(\Delta-2)}\BevenC\,,
\qquad &
\BevenE&=\frac{i(\Delta-\ell -6) (\Delta+\ell -2)}{4
(\Delta-2)}\BevenC\,,\\
\BevenF&=-\frac{3i(\Delta-\ell -6)
(\Delta+\ell -2)}{4 (\Delta-2)}\BevenC\,,\qquad
& \BevenG&=\frac{i(\Delta-\ell -6) (2 \Delta+\ell
-4)}{\Delta-2}\BevenC\,,\\
\BevenH&=\frac{i(2 \Delta-\ell -8)
(\Delta+\ell -2)}{\Delta-2} \BevenC \,,\qquad
& \BevenI&=-\frac{i(\Delta-
\ell -6) (3 \Delta+\ell -8)}{4 (\Delta-2)}\BevenC\,,\\
\BevenJ&=-\frac{i(\Delta-\ell -8) (\Delta-\ell -6)}{4 (\Delta-2)}
\BevenC\,.
\end{aligned}}[WardtBEvenSol]

\subsubsec{Odd \texorpdfstring{$\ell$}{l}}\label{caseBOdd} For general odd
$\ell$ we find the independent structures
\eqn{
t_{\text{B}}^{(\ell\text{ odd})}
(\eta_{i},\etab_{i},U,\Theta\Thetab)=
\frac{ (\eta_3\lnsp\Uup\etab_3)^{\ell-2}}{ U^{4-\Delta+\ell}}\lsp
\left(\sum_{i=1}^{2} \TBOdd{i} \left(A_i +B_i \lsp \xi^2\right) +
\sum_{i=1}^{9} \TBOdd[\llsp\prime\llsp]{i} C_i\right) \,,\qquad
\xi^2 = \lsp\frac{\Theta^2\Thetab^2}{U^2}\,,}[tBOddSpin]
where the structures $\TBOdd{i}$ and $\TBOdd[\llsp\prime\llsp]{i}$ are
defined in Appendix~\ref{appendix:structureEven}.  Similarly to the
even-$\ell$ case, in the lowest component of the three-point function
\ThreePFB for general odd $\ell$ there are two structures, which are related to the $\lambda^{(i)}$ in \JJOnonsusy
 by
\eqn{
\lambda^{(1)}=\lambda^{(2)}=\BoddA-\BoddC\,,\qquad
\lambda^{(3)}=-\lambda^{(4)}=-\BoddC\,.}[BOddZero]

We now need to impose the conservation at the first two points. For
generic $\Delta$ and $\ell$ this gives eleven independent linear
constraints for the thirteen unknowns $\BoddA,\ldots, \BoddM$. Therefore
there are two undetermined coefficients, which we choose to be $\BoddA$ and
$\BoddK$. The result is
\begin{align}
 \BoddC &=\frac{\Delta-\ell -6}{\Delta+\ell
-2}\BoddA\,,\notag\qquad
\BoddB =-\tfrac{1}{4}(\Delta-\ell -6) (\Delta+\ell
-4) \BoddA \,,\notag\\
\BoddD& =-\tfrac{1}{4}(\Delta-\ell -8) (\Delta-\ell
-6) \BoddA \,,\notag\\
\BoddF&=-\frac{i
(\Delta-\ell -8) (\Delta-\ell -6) (\Delta-\ell -2)}{2 (5(\Delta-2)^2 - \ell (\ell+4))} \BoddA
+\frac{(\Delta-\ell -8) (\Delta-\ell -6)}{2
(5(\Delta-2)^2 - \ell (\ell+4))}\BoddK \,,\notag\\
\BoddG&=-\frac{i(\Delta-\ell -6) (\Delta+\ell +2)
(3(\Delta-2)^2+\ell(\ell-4))}{6 (\Delta+\ell -2) (5(\Delta-2)^2
- \ell (\ell+4))} \BoddA
-\frac{(\Delta-\ell-6) (\Delta+\ell +2)}{3 (5(\Delta-2)^2 - \ell (\ell+4))}\BoddK\,,\notag\\
\BoddH&=\frac{2 i \ell  (3 \Delta-\ell -8) (\Delta+\ell
+2)}{3 (5(\Delta-2)^2 - \ell (\ell+4))}\BoddA -\frac{2
(5(\Delta-4)(\Delta-2)-\ell(2\Delta+\ell))}{3 (5(\Delta-2)^2 - \ell (\ell+4))}\BoddK\,,\notag\\
\BoddI&=-\frac{2 i\ell (\Delta-\ell-6)  (\Delta-\ell-2) (3 \Delta+\ell -4)}{3
(\Delta+\ell -2) (5(\Delta-2)^2 - \ell (\ell+4))} \BoddA+
\frac{(\Delta-\ell -6) (5 \Delta+\ell
-10)}{3 (5(\Delta-2)^2 - \ell (\ell+4))}\BoddK\,,\notag\\
\BoddJ&=-\frac{2 i\ell (\Delta-3)
(\Delta-\ell -6)}{5(\Delta-2)^2 - \ell (\ell+4)} \BoddA-\frac{(\Delta-\ell -6) (5 \Delta+\ell -10)}{2 (5(\Delta-2)^2 - \ell (\ell+4))}\BoddK \,,\notag\\
\BoddE&=-\frac{i
(\Delta-\ell -8) (\Delta-\ell -6) (3(\Delta-2)^2+\ell(\ell-4))}{3 (\Delta+\ell -2) (5(\Delta-2)^2 - \ell (\ell+4))} \BoddA-\frac{2(\Delta-\ell -8) (\Delta-\ell
-6)}{3 (5(\Delta-2)^2 - \ell (\ell+4))} \BoddK \,,\notag\\
\BoddL&=\frac{i (\Delta-3)
(\Delta-\ell -6) (\Delta+\ell -2)}{5(\Delta-2)^2 - \ell (\ell+4)}\BoddA +\frac{(\Delta-\ell -6) (3 \Delta+\ell
-4)}{2 (5(\Delta-2)^2 - \ell (\ell+4))}\BoddK\,,\notag\\
\BoddM&=i(\Delta-\ell -6) \BoddA\,.
\label{WardtBOddSol}
\end{align}
\vspace{-\baselineskip}\vspace{8pt}
\newcommand{\WardtBOddSol}{\eqref{WardtBOddSol}\xspace}\\
In the previous cases, studied in Sec.~\ref{caseAsection} and
Sec.~\ref{caseBEven}, imposing conservation at points $z_1$ and $z_2$
automatically implied conservation at point $z_3$ when the dimension of the multiplet $\CO(z_3)$ saturates the
unitarity bound (\UniBoundA or \UniBoundB respectively). In the present
case this is not true: the conservation condition $D_3^{\gamma_1}
\COindellpI{\gamma}=0$ when $\Delta = \ell +4$, or equivalently
\eqn{
\partial_{\eta_3} D_3\,
\vev{\CJ(\eta_1,\etab_1,z_1)\lsp\CJ(\eta_2,\etab_2,z_2)\lsp
\CO_{\ell+2,\ell}(\eta_3,\etab_3,z_3)}\big|_{\Delta=\ell+4}=0\,,
}[Cons3dPointBOddEq]
gives rise to a further non-trivial constraint, which is solved if $\BoddK$ takes the form
\eqn{\BoddK = \frac{2i\ell
(\ell^2+\ell-2)}{(\ell+4)(\ell+5)}\BoddA
+ (\Delta -\ell- 4)\BoddK'\,,}[Cons3rdPointBOdd]
with $\BoddK'$ another arbitrary constant.

\subsec{Case C:~\texorpdfstring{$(\frac12(\ell+4),\frac12\ell)$}
{((l+4)/2,l/2)} operators}
Here we start with
\eqna{
\vev{\CJ(\eta_1,\etab_1,z_1)\lsp\CJ(\eta_2,\etab_2,z_2)\lsp
\CO_{\ell+4,\ell}(\eta_3,\etab_3,z_3)}&\\
&\hspace{-2cm}=
\frac{
  \eta_1\xup_{1\bar{3}}\partial_{\etapb_1}\lsp
  \partial_{\etap_1}\xup_{3\bar{1}}\etab_1\lsp
  \eta_2\xup_{2\bar{3}}\partial_{\etapb_2}\lsp
  \partial_{\etap_2}\xup_{3\bar{2}}\etab_2
}{
  (x_{\bar{3}1}{\lnsp}^2\lsp x_{\bar{1}3}{\lnsp}^2\lsp
  x_{\bar{3}2}{\lnsp}^2\lsp x_{\bar{2}3}{\lnsp}^2)^2
}\lsp
t_{\text{C}}(\etap_{1,2},\eta_3,\etapb_{1,2},\etab_3,U,\Theta\Thetab)\,,
}[ThreePFC]
where $\CO$ has zero R-charge. The unitarity bound is
\eqn{\Delta\geq\ell+6\,,}[UniBoundC]
and homogeneity property now takes the form
\eqn{
t_{\text{C}}
(\eta_{1,2},\kappa\lsp\eta_3,\etab_{1,2},\bar{\kappa}\lsp\etab_3,\lambda\bar{\lambda}\lsp
U,\lambda\llsp\Theta\lsp\llsp \bar{\lambda}\llsp\Thetab) =
(\lambda\bar{\lambda})^{\Delta-6}\kappa^{\ell+4}\bar{\kappa}^\ell\, t_{\text{C}}
(\eta_{i},\etab_{i},U,\Theta\Thetab)\;.
}[]

\subsubsec{Even \texorpdfstring{$\ell$}{l}}
In this case we parametrize the correlator in terms of four tensor structures
\eqn{
t_{\text{C}}^{(\ell\text{ even})}
(\eta_{i},\etab_{i},U,\Theta\Thetab)=
 \frac{ ( \eta_3\lnsp\Uup\etab_3)^{\ell}}{ U^{6-\Delta+\ell}}\lsp
 \left(\TCEven{1} \left(A +B \lsp \xi^2\right) +  \sum_{i=2}^{3} \TCEven{i}
 C_i\right) \,,\qquad
    \xi^2= \lsp\frac{\Theta^2\Thetab^2}{U^2}\,,
  }[tCEvenSpin]
where as usual the quantities $\TCEven{i}$ are defined in Appendix~\ref{appendix:structureEven}. At the lowest order in the Grassaman variables there is only one structure, and the associated parameter is related to the coefficient $\lambda$ in \JJOnonsusy by
\eqn{\lambda = -\CevenA\;.
}[CEvenZero]
Similarly to \WardtBEven we need to require
\eqn{
\partial_{\eta_1}\!\mathcal{D}\,t_{\text{C}}^{(\ell\text{ even})}(\eta_{i},\etab_{i},U,\Theta\Thetab)=0\,,\qquad
\partial_{\etab_1}\!\overbar{\mathcal{D}}\,t_{\text{C}}^{(\ell\text{ even})}(\eta_{i},\etab_{i},U,\Theta\Thetab)=0
\,.}[WardtCEven]
This leads, for generic $\Delta$ and $\ell$, to three linear constraints for the four unknowns $\CevenA,\ldots, \CevenD$. Solving in terms of $\CevenA$ we find
\eqn{\begin{gathered}
\CevenB=-\tfrac{1}{4}(\Delta-\ell -8) (\Delta+\ell -2)\CevenA\,,\qquad
\CevenC=i(\Delta-\ell -8)\CevenA\,,\qquad
\CevenD=i(\Delta+\ell -2)\CevenA\,.
\end{gathered}}[WardtCEvenSol]
Analogously to Sec.~\ref{caseBOdd}, conservation at points
$z_{1,2}$ alone is not sufficient to ensure conservation at point $z_3$
when $\Delta$ saturates the bound \UniBoundC. Instead, the constraint
\eqn{
\partial_{\eta_3}D_3\,
\vev{\CJ(\eta_1,\etab_1,z_1)\lsp\CJ(\eta_2,\etab_2,z_2)\lsp
\CO_{\ell+4,\ell}(\eta_3,\etab_3,z_3)}=0
}[Cons3dPointCEvenEq]
implies that $\CevenA$ must be of the form
\eqn{\CevenA = (\Delta -\ell - 6)\CevenA'\,,}[Cons3rdPointCEven]
where $\CevenA'$ is an arbitrary constant.

\subsubsec{Odd \texorpdfstring{$\ell$}{l}}
Similarly to the even-$\ell$ case we have the four structures
\begin{align}\label{tCOddSpin}
t_{\text{C}}^{(\ell\text{ odd})}
(\eta_{i},\etab_{i},U,\Theta\Thetab)&=
 \frac{ ( \eta_3\lnsp\Uup\etab_3)^{\ell-1}}{ U^{5-\Delta+\ell}}
 \sum_{i=1}^{4} \TCOdd{i} C_i \,,
\end{align}
\vspace{-\baselineskip}\vspace{8pt}
\newcommand{\tCOddSpin}{\eqref{tCOddSpin}\xspace}\\
where the structures $\TCOdd{i}$ are defined in
Appendix~\ref{appendix:structureEven}. This three-point function vanishes
in the limit $\theta_3,\thetab_3 \to 0$, consistently with the fact that
conformal invariance does not allow any structure for a three-point
function of the form $\vev{J(x_1)\lnsp J(x_2)\lnsp
\CO_{\ell+4,\lsp\ell}(x_3)}$ when $\ell$ is odd.

The Ward identities for conservation at points $z_{1,2}$ impose three
constraints on these four constants. Thus, choosing $\CoddA$ as
independent, we obtain
\eqn{\begin{gathered}
\CoddB=-\frac{\Delta-\ell -8}{2(\Delta+\ell +4)}\CoddA\,,\qquad
\CoddC=\frac{3(\Delta-\ell -8)}{2(\Delta+\ell+4)}\lsp \CoddA\,,\qquad
\CoddD=-\frac{6(\Delta-2)}{\Delta+\ell +4}\lsp \CoddA\,.
\end{gathered}}[WardtCOddSol]
As in the even-$\ell$ case conservation at point $z_3$ is not automatic. We need to impose the constraint in \eqref{Cons3dPointCEvenEq} which amounts to requiring $\CoddA$ to be of the form
\eqn{\CoddA = (\Delta-\ell-6)\CoddA'\,.}[Cons3rdPointCOdd]

\newsec{Three-point function coefficients}\label{ThetaThreeExpansion}

In this section we will  set $\theta_{1,2}=\thetab_{1,2} = 0$ and perform
an expansion of the three-point functions presented in Sec.~\ref{Ward} in
$\theta_3,\,\thetab_3$. The results will allow us to extract the OPE
coefficients of the various operators inside the superconformal multiplet
$\CO_{j,\jb}(z)$ in terms of the coefficients $A_i,\,B_i,\,C_i,\,D_i,\,E_i$
defined before. Clearly the zeroth order in $\theta_3,\,\thetab_3$
corresponds to the contribution of the superconformal primary. Similarly
the order $\theta_3\thetab_3$ contains contributions from operators of the
form $(Q\Qb\CO)_{j\pm1,\jb\pm1;p}(x)$, and the order
$\theta_3^2\thetab_3^2$ contains contributions from
$(Q^2\Qb^2\CO)_{j,\jb;p}(x)$, where ``$p$'' stands for primary. However, we
also expect contributions from descendant operators of the schematic form
$(\partial\CO)(x)$ at the first order and of the form
$(\partial^2\CO)(x),\,(\partial Q\Qb \CO)(x)$ at the second order. The
precise way in which this mixing takes place is described
in~\cite{Li:2014gpa}. Following those results we are able to isolate the
contributions of each primary and compute the associated OPE coefficients
in the basis of non-supersymmetric three-point functions adopted
in~\cite{Cuomo:2017wme} and reviewed in Sec.~\ref{Warmup}. The expansion of
\ThreePFAEta in $\theta_3$ can be performed using an extension of the
\emph{Mathematica} package developed in~\cite{Li:2014gpa}. Notice that the
operators $(Q\Qb\CO)_{j\pm1,\jb\pm1;p}(x)$ and $(Q^2\Qb^2\CO)_{j,\jb;p}(x)$
are not normalized in a standard way, but rather according to~\cite[Table
1]{Li:2014gpa}.

In the following we will go through each of the cases presented in
Sec.~\ref{Ward}. In each subsection we explicitly write the non-vanishing
three-point functions between the currents $J$ and the various primary
superconformal descendants in terms of the tensor structures introduced in
Sec.~\ref{Warmup}. In order not to overload the notation we use the same
symbols for the OPE coefficients multiple times---since each subsection
corresponds to a different superprimary, we are confident that this will
not create any confusion. Schematically, we indicate with $\lambda^{(i)}$
the OPE coefficients associated to the superconformal primary,
$\lambda^{(i)}_{\pm\pm}$ or  $\lambda^{(i)}_{\pm\mp}$ those associated to
the superdescendants at order $\theta_3,\thetab_3$, and finally  $\hat
\lambda^{(i)}$ the coefficients of the order $\theta_3^2\thetab_3^2$. The
expression for $\lambda^{(i)}$ were given in the previous section,
Eqs.~\AEvenZero,\coeffsOddZero,\BEvenZero,\BOddZero,\CEvenZero and will not
be repeated.

Here we only report the expression of the $\lambda$'s after imposing the
conservation conditions discussed in the previous section. The coefficients
before the use of conservation conditions can be found in the
\emph{Mathematica} notebook attached to this submission. Because the
conservation condition in superspace \SupercurCons enforces the
conservation of $J$, the $\lambda$'s  satisfy the relations summarized in
Table~\ref{TableConservation}. Thus, we do not show the value of all of
them, but only the independent ones and a few non-trivial relations. We
have checked explicitly that all the constraints of
Table~\ref{TableConservation} are verified.

\subsec{Case A}
\subsubsec{Even \texorpdfstring{$\ell$}{l}}

At order $\theta_3\thetab_3$ we anticipate four different three-point
functions, involving the four different conformal primary operators one can
obtain by acting with $Q\Qb$ on $\CO$~\cite{Li:2014gpa}. We denote them by
$(Q\Qb\CO)_{\ell\pm1,\lsp\ell\pm1;\lsp p}$ and
$(Q\Qb\CO)_{\ell\pm1,\lsp\ell\mp1;\lsp p}$.  In the former case the
resulting primary is still a traceless symmetric operator, but with odd
spin; according to Sec.~\ref{Warmup}, we expect only one tensor structure.
The latter case instead corresponds to a primary with $j-\jb=\pm2$, and we
thus expect four tensor structures parametrized by only one independent
coefficient. In more detail we have
\eqna{\vev{J(\eta_1,\etab_1,x_1)\lsp J(\eta_2,\etab_2,x_2)\lsp
(Q\Qb\CO)_{\ell\pm1,\lsp\ell\pm1;\lsp
p}(\eta_3,\etab_3,x_3)}&=\lambda_{\pm\pm}^{(-)}\lsp
\lsp\mathcal K_{\Delta+1,\ell\pm1,0}\lsp\CS^{(-)}(\eta_{i},\etab_{i},x_{i})\,,
}[pppEven]
where $\lambda_{\pm\pm}^{(-)}$ are coefficients while the remaining
quantities are defined in Sec.~\ref{Warmup}.  The $\pm\pm$ subscript refers
to the addition of unity to the $\ell$ labels of $Q\Qb\CO$ in the left-hand
side, while the sign in the superscript indicates the parity of the
corresponding structure. Similarly, we have the three-point functions
\eqna{
\vev{J(\eta_1,\etab_1,x_1)\lsp J(\eta_2,\etab_2,x_2)\lsp
(Q\Qb\CO)_{\ell+1,\lsp\ell-1;\lsp
p}(\eta_3,\etab_3,x_3)}&= \mathcal K_{\Delta+1,\ell-1,2} \sum_{j=1}^4 \lambda_{+-}^{(j)}\lsp
\CT^{(j)}(\eta_{i},\etab_{i},x_{i})\,,\\
\vev{J(\eta_1,\etab_1,x_1)\lsp J(\eta_2,\etab_2,x_2)\lsp
(Q\Qb\CO)_{\ell-1,\lsp\ell+1;\lsp
p}(\eta_3,\etab_3,x_3)}&= \mathcal K_{\Delta+1,\ell-1,2} \sum_{j=1}^4 \lambda_{-+}^{(j)}\lsp
\overbar\CT^{(j)}(\eta_{i},\etab_{i},x_{i})\, .
}[]
Using \WardtAEvenSol, we obtain the following relations:
\begin{align}
\label{AEvenCoeffs}
\lambda_{++}^{(-)}&=-\frac{2 i (\Delta+\ell ) (\Delta^2-6 \Delta+\ell ^2+8)}{(\ell +1)^2 (3 (\Delta-2)^2-\ell  (\ell +2))}\AevenA -\frac{i (\Delta-\ell -4) (\Delta+\ell -1) (\Delta+\ell )}{(\ell +1)^2 (3 (\Delta-2)^2-\ell  (\ell +2))}\AevenF \,,\notag\\
\lambda_{--}^{(-)}&=
-\frac{2 i  (\ell -1) ( \Delta-\ell -2) (\Delta^2-6 \Delta+\ell ^2+4 \ell +12)}{\ell  (3 (\Delta-2)^2-\ell  (\ell +2))}\AevenA\notag\\
&\hspace{.44cm}
-\frac{i  (\ell +2) (\ell +3) ( \Delta-\ell -2) ( \Delta-\ell -3) (\Delta+\ell -2)}{\ell ^2 (3 (\Delta-2)^2-\ell  (\ell +2))}\AevenF\,,\notag\\
\lambda_{-+}^{(1)}&=-\lambda_{+-}^{(1)}=
\frac{4 i (\Delta-3) (\Delta-\ell -2) (\Delta+\ell )}{(\Delta-1) (3 (\Delta-2)^2-\ell  (\ell +2))}\AevenA -\frac{i (\ell +3) (\Delta-2) (\Delta-\ell -2) (\Delta+\ell )}{\ell (\Delta-1)   (3 (\Delta-2)^2-\ell  (\ell +2))}\AevenF \,,\notag\\
\lambda_{-+}^{(3)}&=\lambda_{+-}^{(3)}=- \frac{\Delta-\ell-4}{2(\ell+1)} \lambda_{+-}^{(1)}\,.
\end{align}
\newcommand{\AEvenCoeffs}{\eqref{AEvenCoeffs}\xspace}

\vspace{-\baselineskip}

\noindent When the unitarity bound \UniBoundA is saturated we get
$\lambda_{\mp\pm}^{(1,3)}=\lambda_{--}^{(-)}=0$. This reflects the
shortening of the multiplet of $\CO$ \cite{Li:2014gpa}.

As expected in \pppEven we do not get any contribution associated with a
parity-even tensor structure.
Also, the relation between $\lambda_{\pm\mp}^{(1)}$ and $\lambda_{\pm\mp}^{(3)}$ is exactly the one expected from Table~\ref{TableConservation} with the proper shifts in the dimension $\Delta$ and spin $\ell$ of the primary.

At order $\theta_3^2\thetab_3^2$ the three-point function \ThreePFAEta
gives rise to a unique three-point function, after subtraction of
three-point functions involving conformal descendants of $\CO$. The final
three-point function involves five parity-even structures,
\eqna{\vev{J(\eta_1,\etab_1,x_1)\lsp J(\eta_2,\etab_2,x_2)\lsp
(Q^2\Qb^2\CO)_{\ell,\lsp\ell;\lsp
p}(\eta_3,\etab_3,x_3)}&=\mathcal K_{\Delta+2,\ell,0} \sum_{j=1}^5\hat{\lambda}^{(j)} \CS^{(j)}(\eta_{i},\etab_{i},x_{i})\,.}[threePFAEvenQsqQbsq]
After imposing the Ward identity constraints \WardtAEvenSol we obtain
\small
\begin{align}
\hat{\lambda}^{(1)}&=
\frac{8(\Delta -2) ( \Delta -\ell -2) (\Delta +\ell )(\Delta ^3-3 \Delta ^2-6 \Delta -\Delta  \ell ^2+9 \ell ^2+2 \Delta ^2 \ell -14 \Delta  \ell +28 \ell +8)}{(\Delta -1)^2 \left(3 (\Delta -2)^2-\ell  (\ell +2)\right)}\AevenA\notag\\
&
+\frac{4 ( \Delta -\ell -2) (\Delta +\ell )
(\Delta ^4-10 \Delta ^3+39 \Delta ^2-82 \Delta -4 \Delta  \ell ^2 +6 \ell ^2-\Delta ^3 \ell +8 \Delta ^2 \ell -35 \Delta  \ell+38 \ell +64)}{(\Delta -1)^2 \left(3 (\Delta -2)^2-\ell  (\ell +2)\right)}\AevenF
\,,\notag\\
\hat{\lambda}^{(4)}&=
-\frac{64\ell (\ell -1) (\Delta-2) (\Delta-\ell -2)  (\Delta+\ell ) }{(\Delta-1)^2 (3 (\Delta-2)^2-\ell  (\ell +2))}\AevenA
\notag\\&\hspace{.45cm}
-\frac{4 (\Delta-\ell -2)(\Delta+\ell )
( 3 \Delta^4-21 \Delta^3-\Delta^2 \ell ^2-2 \Delta^2 \ell +48 \Delta^2-\Delta \ell ^2-2 \Delta \ell -36 \Delta+4 \ell ^2+8 \ell)
}{(\Delta-1)^2 (3 (\Delta-2)^2-\ell  (\ell +2))}\AevenF \,,\notag\\
\hat{\lambda}^{(2)}& =-\hat{\lambda}^{(3)} =-\frac{4  \ell (\Delta -1) \hat{\lambda}^{(1)} -(\Delta -\ell -2) (\Delta +\ell +2)\hat{\lambda}^{(4)}}{2 \Delta ^2+4 \Delta  (\ell +1)-2 \ell    (\ell +4)}\,,\notag\\
\hat{\lambda}^{(5)}& =-\frac{4 \Delta (\Delta -1)  \hat{\lambda}^{(1)} + (\Delta -\ell -2) (\Delta -\ell -4)\hat{\lambda}^{(4)}}{2 \Delta ^2+4 \Delta  (\ell +1)-2 \ell    (\ell +4)}\,,
\end{align}
\normalsize
consistently with the relations of Table~\ref{TableConservation}. All coefficients vanish at the unitarity bound \UniBoundA, as a consequence of the multiplet shortening.

\subsubsec{Odd \texorpdfstring{$\ell$}{l}}

Just as in the even-spin case, at order $\theta_3\thetab_3$ we have four
different three-point functions. Two of these three-point functions involve
five structures (because they contain even-spin operators at
the third point) parametrized by two independent coefficients, while the other two involve four
structures parametrized by one coefficient~\cite{Cuomo:2017wme}. The first three-point function takes the
form
\eqn{\vev{J(\eta_1,\etab_1,x_1)\lsp J(\eta_2,\etab_2,x_2)\lsp
(Q\Qb\CO)_{\ell\pm1,\lsp\ell\pm1;\lsp
p}(\eta_3,\etab_3,x_3)}=\mathcal K_{\Delta+1,\ell\pm1,0}\sum_{i=1}^5\lambda_{\pm\pm}^{(i)}\lsp
\CS^{\llsp(i)}(\eta_{i},\etab_{i},x_{i})\,,}[]
and we also have
\eqna{
\vev{J(\eta_1,\etab_1,x_1)\lsp J(\eta_2,\etab_2,x_2)\lsp
(Q\Qb\CO)_{\ell+1,\lsp\ell-1;\lsp
p}(\eta_3,\etab_3,x_3)}&=\mathcal K_{\Delta+1,\ell-1,2} \sum_{j=1}^4 \lambda_{+-}^{(j)}\lsp
\CT^{(j)}(\eta_{i},\etab_{i},x_{i})\,,\\
\vev{J(\eta_1,\etab_1,x_1)\lsp J(\eta_2,\etab_2,x_2)\lsp
(Q\Qb\CO)_{\ell-1,\lsp\ell+1;\lsp
p}(\eta_3,\etab_3,x_3)}&=\mathcal K_{\Delta+1,\ell-1,2} \sum_{j=1}^4 \lambda_{-+}^{(j)}\lsp
\overbar\CT^{(j)}(\eta_{i},\etab_{i},x_{i})\, .
}[]
The coefficients of these three-point functions are given by, after use of
\WardtAOddSol,

\small
\begin{align}
\lambda_{++}^{(1)}&=\scalemath{0.95}{-\frac{2  (\Delta+\ell ) (3 \Delta^2+2 \Delta \ell -12 \Delta-3 \ell ^2-10 \ell +16)}{(\ell +1)^2 (5 (\Delta-3) (\Delta-1)-3 (\ell -1) (\ell +3))}\AoddA
+\frac{ (\Delta-\ell -4) (\Delta+\ell )}{2 (\ell +1)^2 (5 (\Delta-3) (\Delta-1)-3 (\ell -1) (\ell +3))}\AoddM}
\,,\notag\\
\lambda_{++}^{(4)}&=\scalemath{0.94}{\frac{8  (\Delta-\ell -2) (\Delta+\ell -3) (\Delta+\ell )}{(\ell +1)^2 (5 (\Delta-3) (\Delta-1)-3 (\ell -1) (\ell +3))}\AoddA
-\frac{ (\Delta+\ell ) (3 \Delta^2-6 \Delta-\ell ^2-4 \ell )}{2 (\Delta-2) (\ell +1)^2 (5 (\Delta-3) (\Delta-1)-3 (\ell -1) (\ell +3))}\AoddM}
\,,\notag\\
\lambda_{++}^{(2)}& =-\lambda_{++}^{(3)}=\frac{4 (\Delta -2)(\ell +1)\lambda_{++}^{(1)}- (\Delta -\ell
   -4) (\Delta +\ell +2)\lambda_{++}^{(4)}}{2 (\ell +1) (\ell +9)-2 (\Delta +1)^2-4 (\Delta +1) \ell }\,, \notag\\
\lambda_{++}^{(5)}&=\frac{4 (\Delta -2) (\Delta -1)\lambda_{++}^{(1)}+ (\Delta
   -\ell -4) (\Delta -\ell -6)\lambda_{++}^{(4)}}{2 (\ell +1) (\ell   +9)-2 (\Delta +1)^2-4 (\Delta +1) \ell }\,, \notag\\
\lambda_{--}^{(1)}&=\scalemath{0.9}{-\frac{2 (\Delta-\ell -2) (3 \Delta^2 \ell ^2+\Delta^2 \ell +6 \Delta^2+2 \Delta \ell ^3-2 \Delta \ell ^2+12 \Delta \ell-32 \Delta-3 \ell ^4-19 \ell ^3-28 \ell ^2-28 \ell +48)}{\ell ^2 (5 (\Delta-3) (\Delta-1)-3 (\ell -1) (\ell +3))} \AoddA}
\notag\\
&\hspace{.44cm}
+\frac{ (\ell +3) (\ell +4) (\Delta-\ell -2)^2}{2 \ell ^2 (5 (\Delta-3) (\Delta-1)-3 (\ell -1) (\ell +3))}\AoddM
\,,\notag\\
\lambda_{--}^{(4)}&=\scalemath{0.96}{
\frac{8 (\ell -2) (\ell -1) (\Delta-\ell -2) (\Delta-\ell -5) (\Delta+\ell )}{\ell ^2 (5 (\Delta-3) (\Delta-1)-3 (\ell -1) (\ell +3))}\AoddA
-\frac{ (\ell +3) (\ell +4) (\Delta-\ell -2) (3 \Delta^2-6 \Delta-\ell ^2+4)}{2 \ell ^2(\Delta-2)  (5 (\Delta-3) (\Delta-1)-3 (\ell -1) (\ell +3))}\AoddM}
\,,\notag\\
\lambda_{--}^{(2)}& =-\lambda_{--}^{(3)}=\frac{4(\ell -1) (\Delta -2)\lambda_{--}^{(1)}- (\Delta -\ell
   -2) (\Delta +\ell )\lambda_{--}^{(4)}}{2 \left(\ell ^2-(\Delta -2) (\Delta +2 \ell )-4\right) }\,, \notag\\
\lambda_{--}^{(5)}&=\frac{4 (\Delta -2) (\Delta -1)\lambda_{--}^{(1)}+ (\Delta
   -\ell -2) (\Delta -\ell -4)\lambda_{--}^{(4)}}{2 \left(\ell ^2-(\Delta -2) (\Delta +2 \ell )-4\right)}\,, \notag\\
\lambda_{-+}^{(1)}&=\lambda_{+-}^{(1)}=\frac{4 (\ell -1) (\Delta-4)  (\Delta-\ell -2) (\Delta+\ell )}{\ell  (\ell +1) (5 (\Delta-3) (\Delta-1)-3 (\ell -1) (\ell +3))}\AoddA
\notag\\
&\hspace{1.57cm}
+\frac{(\ell +4)(\Delta-1) (\Delta-\ell -2) (\Delta+\ell )}{2 (\Delta-2) \ell  (\ell +1) (5 (\Delta-3) (\Delta-1)-3 (\ell -1) (\ell +3))}\AoddM
\,,\notag\\
\lambda_{-+}^{(3)}&=-\lambda_{+-}^{(3)}=\frac{ \Delta-\ell -4}{2(\Delta-1) }\lambda_{-+}^{(1)}\,.
\notag\\
\end{align}\normalsize
When the unitarity bound \UniBoundA is saturated,
$\lambda_{\mp\pm}^{(1,2)}=\lambda_{--}^{(j)}=0$, as necessary due
to multiplet shortening.

At order $\theta_3^2\thetab_3^2$ and after subtraction of three-point
functions involving descendants of $\CO$ we are left with a single
three-point function. This involves one independent coefficient $\hat\lambda^{(-)}$,
multiplying the  structure $\CS^{(-)}$, exactly as in \coeffsOddZero but
with $\Delta\to\Delta+2$. We find

\small
\eqna{\hat{\lambda}^{(-)}&=-
\frac{8 i}{\delta_A} (\Delta-2) (\Delta-\ell -2) (\Delta+\ell )\Big(5 \Delta^4-35 \Delta^3-3 \Delta^2 \ell ^2-6 \Delta^2 \ell +84 \Delta^2+13 \Delta \ell ^2+26 \Delta \ell \\[-8pt]
&\hspace{.6cm} -84 \Delta-16 \ell ^2-32 \ell +48\Big)\AoddA
-\frac{4 i}{\delta_A} (\Delta-\ell -2) (\Delta+\ell )(\ell +3) (\ell +4)(\Delta-1)\AoddM
\,,\\
 \delta_A &=(\Delta-2) (\Delta-1)^2 (5 (\Delta-3) (\Delta-1)-3 (\ell -1)
 (\ell +3))\,,
}[]\normalsize
consistently both with the unitarity bound and the fact that there is no
parity-even structure in the three-point function involving two conserved
currents in the first two points when $\ell$ is odd.

\subsec{Case B}
\subsubsec{Even \texorpdfstring{$\ell$}{l}}

At order $\theta_3\thetab_3$ we have four different three-point functions;
one of them is identically zero by conservation (see Table~\ref{TableConservation}), another one involves a traceless symmetric tensor with odd spin, while the other two
have one independent coefficient each. More specifically, we have

\eqna{\vev{J(\eta_1,\etab_1,x_1)\lsp J(\eta_2,\etab_2,x_2)\lsp
(Q\Qb\CO)_{\ell+3,\lsp\ell+1;\lsp
p}(\eta_3,\etab_3,x_3)}&=\mathcal  K_{\Delta+1,\ell+1,2}\sum_{j=1}^4\lambda_{++}^{(j)}\lsp
\CT^{(j)}(\eta_{i},\etab_{i},x_{i})
\,,\\
\vev{J(\eta_1,\etab_1,x_1)\lsp J(\eta_2,\etab_2,x_2)\lsp
(Q\Qb\CO)_{\ell+1,\lsp\ell-1;\lsp
p}(\eta_3,\etab_3,x_3)}&=\mathcal K_{\Delta+1,\ell-1,2}\sum_{j=1}^4\lambda_{--}^{(j)}\lsp
\CT^{(j)}(\eta_{i},\etab_{i},x_{i})
\,,}[ppBEven]
as well as
\eqna{\vev{J(\eta_1,\etab_1,x_1)\lsp J(\eta_2,\etab_2,x_2)\lsp
(Q\Qb\CO)_{\ell+3,\lsp\ell-1;\lsp
p}(\eta_3,\etab_3,x_3)}&=0\,,}[pmBEven]
and
\eqna{\vev{J(\eta_1,\etab_1,x_1)\lsp J(\eta_2,\etab_2,x_2)\lsp
(Q\Qb\CO)_{\ell+1,\lsp\ell+1;\lsp
p}(\eta_3,\etab_3,x_3)}&=\mathcal K_{\Delta+1,\ell+1,0}  \lambda_{-+}^{(-)}\lsp
\CS^{(-)}(\eta_{i},\etab_{i},x_{i})\,.}[mpBEven]

Using Eq.~\WardtBEvenSol, the above coefficients are given by
\begin{align}
\lambda_{++}^{(1)}&=-\frac{i(\Delta+\ell) (\Delta+\ell+2)}{(\ell+1)
(\Delta-2) (\Delta+\ell +1)}\BevenC\,,
&\lambda_{++}^{(3)}&=-\frac{\Delta-\ell-6}{2 (\ell +3)}\lambda_{++}^{(1)}\,,\notag\\
\lambda_{--}^{(1)}&=-\frac{i(\ell +1) (\ell +3) (\ell +4) (\Delta-\ell -4)
(\Delta-\ell -2)}{\ell  (\ell +2) (\Delta-2) (\Delta-\ell -3)} \BevenC \,,
&\lambda_{--}^{(3)}&=-\frac{ \Delta-\ell -4 }{2 (\ell +1)} \lambda_{--}^{(1)} \, ,\notag\\
\lambda_{-+}^{(-)}&=\frac{i (\ell +4)(\Delta-3) (\Delta-\ell -4) (\Delta+\ell
)}{ (\ell +1) (\ell +2)(\Delta-2)^2} \BevenC \,.\notag\\
\end{align}
When the unitarity bound \UniBoundB is saturated, $\lambda_{-+}^{(-)}=\lambda_{--}^{(1,2)}=0$, as required by multiplet shortening.\par
At order $\theta_3^2\thetab_3^2$ we expect four tensor structures parametrized by one independent coefficient. The structures are the same as the
ones in \BEvenZero but with $\Delta \to \Delta +2$. If we denote
\eqna{\vev{J(\eta_1,\etab_1,x_1)\lsp J(\eta_2,\etab_2,x_2)\lsp
(Q^2\Qb^2\CO)_{\ell+2,\lsp\ell;\lsp
p}(\eta_3,\etab_3,x_3)}&=\mathcal K_{\Delta+2,\ell,2}\sum_{j=1}^4 \hat\lambda^{(j)}\lsp\CT^{(j)}(\eta_{i},\etab_{i},x_{i})\,,}[threePFBEvenQsqQbsq]
then, after using \WardtBEvenSol, we obtain
\begin{align}
 \hat\lambda^{(1)}&=-\frac{4 (\Delta-3) (\Delta-\ell -4) (\Delta-\ell -2)
(\Delta+\ell ) (\Delta+\ell +2)}{(\Delta-2) (\Delta-\ell -3) (\Delta+\ell +1)}\BevenC \,,
& \hat\lambda^{(3)}&=-\frac{ \Delta-\ell -4 }{2\Delta}  \hat\lambda^{(1)}\,.
\end{align}
As expected all $ \hat\lambda^{(j)}$ go to zero when the bound \UniBoundB is saturated.

\subsubsec{Odd \texorpdfstring{$\ell$}{l}}
At order $\theta_3\thetab_3$ we have four different three-point functions.
One of them is parametrized by only one coefficient, another is the three-point function of
a symmetric traceless operator with even spin and contains therefore two independent coefficients, while the other two contain one coefficient each. We have
\eqna{\vev{J(\eta_1,\etab_1,x_1)\lsp J(\eta_2,\etab_2,x_2)\lsp
(Q\Qb\CO)_{\ell+3,\lsp\ell+1;\lsp
p}(\eta_3,\etab_3,x_3)}&=\mathcal K_{\Delta+1,\ell+1,2}\sum_{j=1}^4\lambda_{++}^{(j)}\lsp
\CT^{(j)}(\eta_{i},\etab_{i},x_{i})
\,,\\
\vev{J(\eta_1,\etab_1,x_1)\lsp J(\eta_2,\etab_2,x_2)\lsp
(Q\Qb\CO)_{\ell+1,\lsp\ell-1;\lsp
p}(\eta_3,\etab_3,x_3)}&=\mathcal K_{\Delta+1,\ell-1,2}\sum_{j=1}^4\lambda_{--}^{(j)}\lsp
\CT^{(j)}(\eta_{i},\etab_{i},x_{i})
\,,}[ppBOdd]
and
\eqna{\vev{J(\eta_1,\etab_1,x_1)\lsp J(\eta_2,\etab_2,x_2)\lsp
(Q\Qb\CO)_{\ell+3,\lsp\ell-1;\lsp p}(\eta_3,\etab_3,x_3)}&=\lambda_{+-}\lsp
\mathcal K_{\Delta+1,\ell-1,4}
\mathcal R(\eta_{i},\etab_{i},x_{i})\,,}[pmBOdd]
while the symmetric traceless one is
\eqn{\vev{J(\eta_1,\etab_1,x_1)\lsp J(\eta_2,\etab_2,x_2)\lsp
(Q\Qb\CO)_{\ell+1,\lsp\ell+1;\lsp
p}(\eta_3,\etab_3,x_3)}=\mathcal K_{\Delta+1,\ell+1,0} \sum_{i=1}^5\lambda_{-+}^{(i)}\lsp
\CS^{\llsp(i)}(\eta_{i},\etab_{i},x_{i})\,,}[mpBOdd]

With the use of \WardtBOddSol and \eqref{Cons3rdPointBOdd} we obtain
\small
\begin{align}
\lambda_{++}^{(1)}=&-
\frac{2 i}{\delta_2} (\Delta-1) (\Delta+\ell +2)\Big(5 \ell ^4+4 \Delta \ell ^3+19\ell ^3+3\Delta^2 \ell ^2-14 \Delta \ell ^2+58 \ell ^2+27 \Delta^2\ell-170 \Delta \ell \notag\\[-8pt]
&+248 \ell +60 \Delta^2-360 \Delta+480\Big)\BoddA -\frac{4}{\delta_1} (\Delta-1)
(\Delta-\ell -4) (\Delta+\ell +2)\BoddK'
\,,\notag\\
\lambda_{++}^{(3)} =& -\frac{\Delta-\ell-6}{2(\Delta-1)}\lambda_{++}^{(1)}
\,,\notag\\
\lambda_{--}^{(1)}=&-\frac{2 i}{\delta_{4}}(\ell -1) (\Delta-1) (\Delta-\ell -4) (\Delta-\ell -2)  (3\Delta^2+4 \ell  \Delta-16 \Delta+5 \ell ^2+4 \ell +36) \BoddA\notag\\[-8pt]
&\quad\quad-\frac{4}{\ell\delta_{3}}  (\ell +4) (\ell +5)(\Delta-1) (\Delta-\ell -4)
(\Delta-\ell -2)\BoddK'
\, ,\notag\\
\lambda_{--}^{(3)} =& -\frac{\Delta-\ell-4}{2(\Delta-1)}\lambda_{--}^{(1)}
\, ,\notag\\
\lambda_{+-}=&\frac{4 i}{\Delta\delta_6} (\Delta-\ell -4) (\Delta-\ell -2)(\Delta+\ell +2) (3 \ell\Delta^2+12 \Delta^2+2 \ell ^2 \Delta-4 \ell\Delta -58 \Delta-3 \ell^2-12 \ell +60) \BoddA\notag\\[-8pt]
&\quad\quad-\frac{2}{\ell\delta_5} (\ell +5)(\Delta-\ell-2) (\Delta-\ell-4) (\Delta+\ell +2) \BoddK'
\,,
\notag\\
\lambda_{-+}^{(1)}=&\frac{2 i}{\delta_8}(\Delta-\ell-4) \Big(3\Delta \ell ^5-12 \ell ^5-5 \Delta^2 \ell ^4+59 \Delta \ell ^4-176\ell ^4-19 \Delta^3 \ell ^3+135 \Delta^2 \ell ^3-132 \Delta \ell^3 -496 \ell ^3\notag \\[-8pt]
&-3 \Delta^4 \ell ^2-115 \Delta^3 \ell ^2+1056 \Delta^2\ell ^2-2412 \Delta \ell ^2+800 \ell ^2 -27 \Delta^4 \ell -136 \Delta^3 \ell +2324 \Delta^2 \ell -6608 \Delta \ell\notag \\[-8pt]
& +4864 \ell -60 \Delta^4+240 \Delta^3+720 \Delta^2-3840 \Delta+3840\Big) \BoddA -\frac{4}{\delta_7}(\ell +5) (\Delta-2) (\Delta-\ell -4)^2 \BoddK'
\,,
\notag\\
\lambda_{-+}^{(4)}=&-\frac{4 i \ell}{\delta_8} (\Delta-\ell -4) \Big(3 \ell  \Delta^4+12 \Delta^4-3 \ell^2 \Delta^3-21 \ell  \Delta^3-96 \Delta^3-3 \ell ^3 \Delta^2+45 \ell^2 \Delta^2+168 \ell  \Delta^2\notag \\[-8pt]
&\quad +360 \Delta^2-\ell ^4 \Delta+13 \ell ^3\Delta-128 \ell ^2 \Delta-532 \ell  \Delta-672 \Delta-\ell ^4 -44 \ell ^3+4 \ell ^2+416 \ell +480\Big) \BoddA\notag\\[-8pt]
&\quad+\frac{4}{\delta_7} (\ell +5) (\Delta-\ell -4) (3 \Delta(\Delta-2)
-\ell(\ell+4)) \BoddK'
\,,
\notag\\
\lambda_{-+}^{(2)} =&\frac{4 (\ell +1)(\Delta -2)\lambda_{-+}^{(1)}- (\Delta -\ell
   -4) (\Delta +\ell +2)\lambda_{-+}^{(4)}}{2 (\ell +1) (\ell +9)-2 (\Delta +1)^2-4 (\Delta +1) \ell }\,, \notag\\
\lambda_{-+}^{(5)}=&\frac{4 (\Delta -2) (\Delta -1)\lambda_{-+}^{(1)}+ (\Delta
   -\ell -4) (\Delta -\ell -6)\lambda_{-+}^{(4)}}{2 (\ell +1) (\ell   +9)-2 (\Delta +1)^2-4 (\Delta +1) \ell }\,, \notag\\
\end{align}
\normalsize
where we have defined
\begin{align}
\delta_1 &=3 (\ell +1) (\ell +3) (5(\Delta-2)^2-\ell(\ell+4)) \,,
&\delta_2&=(\ell +4) (\ell +5) (\Delta+\ell +1) \delta_1\,,\notag\\
\delta_3 &=3 (\ell +2) (5(\Delta-2)^2-\ell(\ell+4))\,,
&\delta_4 &=(\Delta-\ell-3)(\Delta+\ell -2) \delta_3\,, \notag\\
\delta_5 &=3 (\ell +3) (5(\Delta-2)^2-\ell(\ell+4))\,,
&\delta_6 &=(\ell +4) (\Delta+\ell-2)\delta_5\notag\,,\\
\delta_7 &=3 (\ell +1) (\ell +2) (5(\Delta-2)^2-\ell(\ell+4))\,,
&\delta_8&=(\ell +4) (\Delta-2) (\Delta+\ell -2) \delta_7\,.
\end{align}
When the unitarity bound \UniBoundB is saturated,
$\lambda_{-+}^{(1,\ldots,4)}=\lambda_{+-}=\lambda_{--}^{(1,2)}=0$, as
required by multiplet shortening.

At order $\theta_3^2\thetab_3^2$ we expect two structures parametrized by one coefficient. These structures are the same as the ones in \BOddZero but with $\Delta \to \Delta +2$.
More specifically, we have
\eqna{\vev{J(\eta_1,\etab_1,x_1)\lsp J(\eta_2,\etab_2,x_2)\lsp
(Q^2\Qb^2\CO)_{\ell+2,\lsp\ell;\lsp
p}(\eta_3,\etab_3,x_3)}&=\mathcal K_{\Delta+2,\ell,2}\sum_{j=1}^4\hat \lambda^{(j)}\lsp
\CT^{(j)}(\eta_{i},\etab_{i},x_{i})\,.}[threePFBOddQsqQbsq]
After using \WardtBOddSol and \eqref{Cons3rdPointBOdd} we obtain
\small
\begin{align}
\hat{\lambda}^{(1)}=&-
\frac{8}{\delta_{10}}(\ell +2) (\Delta-\ell -4) (\Delta-\ell -2)(\Delta+\ell +2) \Big(15 \Delta^5+15 \Delta^4 \ell -135 \Delta^4-3\Delta^3 \ell ^2 -147 \Delta^3 \ell\notag\\[-8pt]
&+390 \Delta^3 -3 \Delta^2 \ell^3  +27 \Delta^2 \ell ^2+546 \Delta^2 \ell -300 \Delta^2+31 \Delta \ell^3-2 \Delta \ell ^2-884 \Delta\ell -360 \Delta-46 \ell ^3-112 \ell^2\notag\\[-8pt]
&+488 \ell +480\Big)\BoddA  -\frac{32 i}{\delta_9} (\ell +2) (\ell +4) (\ell +5)(\Delta-1)
(\Delta-\ell -4) (\Delta-\ell -2) (\Delta+\ell +2) \BoddK'
\,,
\notag\\[8pt]
\hat{\lambda}^{(3)} =& -\frac{\Delta-\ell-4}{2(\ell+2)}\hat{\lambda}^{(1)}
\end{align}
\normalsize
where we defined the denominators
\eqna{
\delta_{9}&=3 \Delta (\Delta-2) (\Delta-\ell -3)(\Delta+\ell +1) (5(\Delta-2)^2-\ell(\ell+4))\,,\\
\delta_{10}&= (\Delta+\ell-2)\delta_9\,.
}[]
Consistently with multiplet shortening $\hat{\lambda}^{(j) }=
0$ at the unitarity bound \UniBoundB.

\subsec{Case C}
\subsubsec{Even \texorpdfstring{$\ell$}{l}}

At order $\theta_3\thetab_3$ only one three-point function is non-zero, namely
\eqna{\vev{J(\eta_1,\etab_1,x_1)\lsp J(\eta_2,\etab_2,x_2)\lsp
(Q\Qb\CO)_{\ell+3,\lsp\ell+1;\lsp
p}(\eta_3,\etab_3,x_3)}&=\mathcal K_{\Delta+1,\ell+1,2}\sum_{j=1}^4 \lambda^{(j)}_{-+}\lsp
\CT^{(j)}(\eta_{i},\etab_{i},x_{i})\,.}[mpCEven]
With the use of \WardtCEvenSol and \eqref{Cons3rdPointCEven} we find
\eqna{\lambda_{-+}^{(1)}&=\frac{2i(\ell +3) (\ell +5) (\Delta-4)
(\Delta-\ell -6)}{ (\ell +1) (\ell +4)(\Delta-3)}\CevenA'\,,
\qquad \lambda_{-+}^{(3)}=-\frac{ \Delta-\ell -6}{2(\ell +3)}\lambda_{-+}^{(1)}\, .
}[]
As expected from multiplet shortening, both coefficients vanish when
$\Delta$ saturates \UniBoundC.

At order $\theta_3^2\thetab_3^2$ we have only one structure, namely the
same one multiplied by $\lambda$ in \CEvenZero with $\Delta\to\Delta+2$,
\eqna{\vev{J(\eta_1,\etab_1,x_1)\lsp J(\eta_2,\etab_2,x_2)\lsp
(Q^2\Qb^2\CO)_{\ell+4,\lsp\ell;\lsp
p}(\eta_3,\etab_3,x_3)}&=\hat{\lambda}\lsp \mathcal K_{\Delta+2,\ell,4}
\mathcal R (\eta_{i},\etab_{i},x_{i})\,.}[threePFCEvenQsqQbsq]
The coefficient is determined to be
\begin{align}
\hat{\lambda}&=\frac{4(\Delta-4) (\Delta-\ell -6) (\Delta-\ell -2)
(\Delta+\ell +4)}{\Delta+1}\CevenA'\, ,
\end{align}
which vanishes for $\Delta$ saturating \UniBoundC as needed.

\subsubsec{Odd \texorpdfstring{$\ell$}{l}}
We have three non-zero three-point functions with one independent coefficient each, i.e.\
\eqn{\begin{aligned}
\vev{J(\eta_1,\etab_1,x_1)\lsp J(\eta_2,\etab_2,x_2)\lsp
(Q\Qb\CO)_{\ell+5,\lsp\ell+1;\lsp
p}(\eta_3,\etab_3,x_3)}&=\lambda_{++}\lsp \mathcal K_{\Delta+1,\ell+1,4}
\mathcal R (\eta_{i},\etab_{i},x_{i})\,,\notag\\
\vev{J(\eta_1,\etab_1,x_1)\lsp J(\eta_2,\etab_2,x_2)\lsp
(Q\Qb\CO)_{\ell+3,\lsp\ell-1;\lsp
p}(\eta_3,\etab_3,x_3)}&=\lambda_{--}\lsp \mathcal K_{\Delta+1,\ell-1,4}
\mathcal R (\eta_{i},\etab_{i},x_{i})\,,
\end{aligned}
}[ppCOdd]
and
\eqna{\vev{J(\eta_1,\etab_1,x_1)\lsp J(\eta_2,\etab_2,x_2)\lsp
(Q\Qb\CO)_{\ell+3,\lsp\ell+1;\lsp
p}(\eta_3,\etab_3,x_3)}&=\mathcal K_{\Delta+1,\ell+1,2} \sum_{j=1}^4 \lambda_{-+}^{(j)}\lsp
\CT^{\lsp (j)}(\eta_{i},\etab_{i},x_{i})\,.}[mpCOdd]
With the use of \WardtCOddSol and \eqref{Cons3rdPointCOdd} we obtain
\eqn{\begin{aligned}
\lambda_{++}&=-\frac{\Delta-\ell -6}{(\ell +1) (\ell +5)}\CoddA'\,,\\
\lambda_{--}&=-\frac{(\ell +5)(\ell+6) (\Delta-\ell -6) (\Delta-\ell
-2) }{\ell(\ell +4)(\Delta+\ell+4)}\CoddA'\, ,\\
\lambda_{-+}^{(1)}&=-\frac{2(\ell+6)(\Delta-1)(\Delta-\ell-6)}
{(\ell+1)(\ell+4)(\Delta+\ell+4)}\CoddA'
\,,\\
 \lambda_{-+}^{(3)}&=- \frac{\Delta-\ell-6}{2(\Delta-1)} \lambda_{-+}^{(1)}\,.
\end{aligned}}[]
As expected $\lambda^{(j)}_{-+} = \lambda_{--}=0$ at the unitarity bound \UniBoundC, due to multiplet shortening.

The $\theta_3^2\thetab_3^2$ order is purely a descendant, consistently with
the fact that there is no three-point function at the lowest order, i.e.\
one has
\eqna{\vev{J(\eta_1,\etab_1,x_1)\lsp J(\eta_2,\etab_2,x_2)\lsp
(Q^2\Qb^2\CO)_{\ell+4,\lsp\ell;\lsp
p}(\eta_3,\etab_3,x_3)}&=0\,.}[threePFCPddQsqQbsq]
We checked that this is indeed the case.

\newsec{Discussion}

In this work we have studied the constraints imposed by $\mathcal N=1$
superconformal  symmmetry on the three-point function involving two
Ferrara--Zumino supermultiplets $\mathcal J_{\alpha\alphad}$ and a third
supermultiplet $\mathcal O_{j,\jb}$.  We started from the  most general
parametrization of such a correlator in superspace and we imposed the
shortening conditions $D^\alpha \mathcal
J_{\alpha\alphad}=\overbar{D}^\alphad \mathcal J_{\alpha\alphad}=0$.
Similarly to the non supersymmetric case, reviewed in Sec.~\ref{Warmup},
these contraints give rise to linear equations among the coefficients of
the superspace tensor structures. We found that non-trivial solutions exist
only when the superconformal primary contained in $\mathcal O_{j,\jb}$ has
a non-vanishing three-point function with two the R-currents
$J_\mu$.\footnote{To be precise, when $j=\jb\pm4$ and $j$ is odd the
non-supersymmetric case vanishes while the supersymmetric one is allowed.}
This is equivalent to requiring that $\mathcal O_{j,\jb}$ is neutral under
the R-symmetry and satisfies $j-\jb=0,\pm2,\pm4$. We also give a group
theoretical counting of the number of independent superspace tersor
structures expected for this correlator.

The results of Sec.~\ref{Ward} are completely general and in principle
contain all the information needed in order to extract the three-point
function between any two conserved currents and a third operator.  A
striking consequence of our results is that the OPE of any two conserved
currents in 4d $\mathcal N=1$ SCFTs contains at most the same
superconformal primaries entering the OPE $J_\mu\times J_\nu$. This
represents a big difference compared with the non-supersymmertric case
where, for instance, the OPE of two stress-energy tensors contains more
general representations.  For this reason we expect that a bootstrap study
of the correlation function involving four copies of the R-current $J_\mu$
will be able to capture interesting features of $\mathcal N=1$ theories.

This last point was the main motivation for the present work: setting the
stage for a numerical bootstrap study of the correlation function $\langle
J_\mu J_\nu J_\rho J_\sigma \rangle$. A key ingredient needed for this
analysis is the superconformal block decomposition of the correlator.
Introducing a proper basis of tensor structures $\mathbb T^{(i)}_4$ we can  schematically write
\eqn{\langle J_\mu J_\nu J_\rho J_\sigma \rangle \propto \sum_{\mathcal O} \sum_{s=1}^{n_4} W_\mathcal{O}^{(i)} \mathbb T^{(i)}_4\,,
}[]
where we have omitted a kinematic prefactor, and the partial waves
$W_\mathcal{O}^{(i)}$ represent the contribution of an entire
superconformal multiplet to the four-point function,
\eqn{W_\mathcal{O}^{(i)} = \sum_{O'=\mathcal O, Q\overbar Q\mathcal{O},\ldots} \sum_{a,b=1}^{n_{O'}} \lambda_{JJ O'}^a  \lambda_{\overbar{O}' JJ}^b g^{(a,b;\llsp i)}_{\Delta_{O'},\ell_{ O'}}\,.
 }[partialWave]
In the above expression $g^{(a,b;\llsp i)}_{\Delta_{O'},\ell_{ O'}}$ are
the non-supersymmetric conformal blocks for spinning correlators computed
in \cite{Echeverri:2016dun, Echeverri:2015rwa} and the
$\lambda$'s are the three-point function coefficients between two currents
and the various components of the supermultiplet $\mathcal O$; because of
supersymmetry they are all determined in terms of the superconformal
primary ones.  The main result of this work is the exact form of these
relations. This was achieved in Sec.~\ref{ThetaThreeExpansion}. Notice that
the coefficients presented there do not correspond exactly to the ones in
\partialWave since the superconformal descendants do not have a standard
normalization. Thus, one has to divide by their norms, which have been
already computed~\cite{Li:2014gpa}.

As anticipated, the next step is to bootstrap numerically the correlation
function of four R-currents, along the same lines as the 3d global symmetry
current bootstrap~\cite{Dymarsky:2017xzb} and the 3d stress-energy tensor
bootstrap~\cite{Dymarsky:2017yzx}. In particular one has to remove the
redundant crossing symmetry conditions by properly taking into account the
conservation constraint. This step could be more involved in the
supersymmetric case.  Given the universal nature of the Ferrara--Zumino
multiplet, this study will open a window on all local 4d $\mathcal N=1$
SCFTs, potentially leading to discovering new ones and hopefully shedding
light on the putative minimal SCFT studied in~\cite{Poland:2011ey,
Poland:2015mta, Li:2017ddj}.

The three-point functions computed in this work can also be used in
combination with lightcone bootstrap techniques~\cite{Fitzpatrick:2012yx,
Komargodski:2012ek} and the recent OPE inversion formula
\cite{Caron-Huot:2017vep, Simmons-Duffin:2017nub} or the methods of
\cite{Alday:2016njk} to study the behavior of Regge trajectories in
supersymmetry.

Finally, starting from our results in superspace, with a bit more technical
effort it is possible to compute the three-point functions of other
components of $\mathcal J$, such as the stress-energy tensor. Although at
the level of three-point functions one does not get any additional
information, it is not excluded that the crossing constraints coming from
their four-point functions would impose independent restrictions. It would
be interesting to investigate this direction.

\ack{We would like to thank Denis Karateev, Petr Kravchuk and Madalena
Lemos for interesting discussions.  AS would like to thank EPFL and IHES
for hospitality throughout and during the final stages of this work,
respectively. AM and AV are supported by the Swiss National Science
Foundation under grant no.\ PP00P2-163670.}

\begin{appendices}
\newsec{Tensor structures in spinor formalism}
\label{appendix:structureEven}

Here we list the tensor structures appearing in the three-point functions of this paper. Let us introduce the following useful notation:
\eqn{
\begin{gathered}
  \Uc{i}{\jmath} = \frac{\eta_i\lnsp\Uup\etab_j}{|U|} \,,\quad  \Uc{\Theta}{\Theta} = \frac{\Theta\Uup\Thetab}{U^2}\,, \quad   \Ec{i}{j} = \eta_i \eta_j\,, \quad \Ecb{\imath}{\jmath} = \bar{\eta}_i \bar{\eta}_j\,,\\
  \Ec{\Theta}{j} = \frac{\Theta \eta_j}{|U|^{1/2}} \,,\quad
    \Ecb{\Theta}{\jmath} = \frac{\Thetab \etab_j}{|U|^{1/2}} \,,\quad
    \Uc{j}{\Theta} = \frac{\eta_i\lnsp\Uup\Thetab}{|U|^{3/2}}\,, \quad \Uc{\Theta}{\jmath} = \frac{\Theta\lnsp\Uup\etab_j}{|U|^{3/2}}\,.
\end{gathered}}[shortcuts]
Let us start with case \ref{caseA} for $\ell$ even. In \tAEvenSpinEta we have the following tensor structures:
\begin{align}\label{AEvenStructures}
\TAEven{1}&=\Ec{1}{2}\Ecb{1}{2}\Uc{3}{3}^2\,, \nonumber\\
\TAEven{2} &=\lsp  \Ec{1}{3}\Ec{2}{3}\Ecb{1}{3}\Ecb{2}{3}  \,,\nonumber\\
\TAEven{3} &=(\lnsp\Uc{1}{1} \Uc{2}{2}+\Uc{1}{2}\Uc{2}{1}\llsp) \Uc{3}{3}^2\,,\nonumber\\
\TAEven{4} &=    \left(\Ec{1}{3} (\Ecb{1}{3} \Uc{2}{2} +\lsp \Ecb{2}{3} \Uc{2}{1}\lsp ) +  \Ec{2}{3} \lsp( \Ecb{1}{3} \Uc{1}{2} +\lsp \Ecb{2}{3} \Uc{1}{1} \lsp )\right)\lsp \Uc{3}{3}\,, \nonumber\\
\TAEven[\llsp\prime\llsp]{1} &=i   \left(\Ec{1}{2} ( \Uc{\Theta}{1} \Ecb{\Theta}{2} + \lsp \Uc{\Theta}{2} \Ecb{\Theta}{1} ) - \Ecb{1}{2} ( \Ec{\Theta}{1} \Uc{2}{\Theta} + \lsp \Ec{\Theta}{2} \Uc{1}{\Theta} )\right)\lsp \Uc{3}{3}^2 \,, \nonumber\\
\TAEven[\llsp\prime\llsp]{2} &=  i \left(\Ec{1}{2} ( \Ecb{1}{3}\Uc{3}{2} + \lsp \Ecb{2}{3} \Uc{3}{1} ) - \Ecb{1}{2} (  \Ec{1}{3}\Uc{2}{3} + \lsp \Ec{2}{3} \Uc{1}{3} )  \right)\Ec{\Theta}{3}\Ecb{\Theta}{3} \,, \nonumber\\
\TAEven[\llsp\prime\llsp]{3} &= \left(\Ec{1}{2} \lsp ( \Ecb{1}{3}\Uc{3}{1} + \lsp \Ecb{2}{3} \Uc{3}{1} ) + \Ecb{1}{2} (  \Ec{1}{3}\Uc{2}{3} + \lsp \Ec{2}{3} \Uc{1}{3} )  \right)\Ec{\Theta}{3}\Ecb{\Theta}{3} \,, \nonumber\\
\TAEven[\llsp\prime\llsp]{4}&= i \left(\Ec{1}{2} \Ec{\Theta}{3}(\lnsp  \Ecb{\Theta}{1}\Ecb{2}{3} + \lsp  \Ecb{\Theta}{2}\Ecb{1}{3} ) - \Ecb{1}{2}(  \Ec{\Theta}{2}\Ec{1}{3} + \lsp  \Ec{\Theta}{1}\Ec{2}{3} )  \Ecb{\Theta}{3} \lsp\right) \Uc{3}{3} \,, \nonumber\\  \nonumber
\TAEven[\llsp\prime\llsp]{5} &=i \left(\Ec{1}{2} (  \Ecb{1}{3}\Uc{3}{1} + \lsp \Ecb{2}{3} \Uc{3}{1} ) - \Ecb{1}{2} (  \Ec{1}{3}\Uc{2}{3} + \lsp \Ec{2}{3} \Uc{1}{3} )  \right) \Uc{3}{3}\Uc{\Theta}{\Theta} \,, \nonumber\\
\TAEven[\llsp\prime\llsp]{6} &= \Uc{3}{3} \lsp\big( \Ecb{1}{2} (\Ec{2}{3}\lsp \Ec{\Theta}{1}\lsp \Ecb{\Theta}{3}+ \Ec{1}{3}\lsp \Ec{\Theta}{2} \lsp\Ecb{\Theta}{3})
 +\Ec{1}{2}(\Ecb{2}{3} \lsp\Ec{\Theta}{3}\lsp\Ecb{\Theta}{1}+\Ecb{1}{3}\lsp \Ec{\Theta}{3}\lsp\Ecb{\Theta}{2})\big)\nonumber\\
& \hspace{.42cm} -2\llsp \Ec{1}{2}\lsp\Ecb{1}{3}\lsp\Ecb{2}{3}\lsp \Ec{\Theta}{3}\lsp\Uc{3}{\Theta}
-2\llsp \Ecb{1}{2}\lsp\Ec{1}{3}\lsp\Ec{2}{3}\lsp\Uc{\Theta}{3}\lsp\Ecb{\Theta}{3} \,,\nonumber\\
\TAEven[\llsp\prime\llsp]{7} &=  \left(\Ec{1}{2} (\Ecb{2}{3} \lsp \Uc{\Theta}{1}+\Ecb{1}{3} \lsp \Uc{\Theta}{2}) \Uc{3}{\Theta}  + \Ecb{1}{2}\Uc{\Theta}{3} (\Ec{2}{3} \lsp \Uc{1}{\Theta} +\Ec{1}{3} \lsp \Uc{2}{\Theta})  )\right)\Uc{3}{3}\,, \nonumber\\
\TAEven[\llsp\prime\llsp]{8}  &=   \left(\Ec{1}{2}   \Uc{3}{1}\Uc{3}{2} \Uc{\Theta}{3} \Ecb{\Theta}{3} +\Ecb{1}{2}   \Uc{1}{3}\Uc{2}{3} \Ec{\Theta}{3} \Uc{3}{\Theta} )  \right) \,.
\end{align}
\vspace{-\baselineskip}\vspace{8pt}
\newcommand{\AEvenStructures}{\eqref{AEvenStructures}\xspace}\\
If $\ell = 0$ only the structures $\TAEven{1},\,\TAEven{3}$ and $\TAEven[\llsp\prime\llsp]{1}$ are present.
The structures for the odd-spin case appearing in \tAOddSpinEta read
\begin{align}\label{AOddStructures}
\TAOdd{1}&= i\left(\Ec{1}{2}\lsp(\Ecb{1}{3} \Uc{3}{2} + \Ecb{2}{3} \Uc{3}{1})-
            \lsp \Ecb{1}{2}\lsp(\Ec{1}{3} \Uc{2}{3} + \Ec{2}{3} \Uc{1}{3})\right)\lsp\Uc{3}{3}^2\,, \nonumber\\
\TAOdd{2} &=\left(\Ec{1}{2}\lsp(\Ecb{1}{3} \Uc{3}{2} + \Ecb{2}{3} \Uc{3}{1})+
            \lsp \Ecb{1}{2}\lsp(\Ec{1}{3} \Uc{2}{3} + \Ec{2}{3} \Uc{1}{3})\right)\lsp \Uc{3}{3}^2\,,\nonumber\\
\TAOdd[\llsp\prime\llsp]{1} &= \Ec{1}{3}\Ec{2}{3}\Ecb{1}{3}\Ecb{2}{3}  \,,\nonumber\\
\TAOdd[\llsp\prime\llsp]{2} &= \left(\Uc{1}{1} \Uc{2}{2} + \Uc{1}{2}\Uc{2}{1}\lsp \right)\Uc{3}{3}^2\,,\nonumber\\
\TAOdd[\llsp\prime\llsp]{3} &= \Ec{1}{2}\Ecb{1}{2} \Uc{3}{3}^2 \,,\nonumber\\
\TAOdd[\llsp\prime\prime\llsp]{1} &=\left(
            \Ec{1}{3} \Ecb{1}{3} \Ec{\Theta}{2}\Ecb{\Theta}{2}+
            \Ec{1}{3} \Ecb{2}{3} \Ec{\Theta}{2}\Ecb{\Theta}{1}+
            \Ec{2}{3} \Ecb{1}{3} \Ec{\Theta}{1}\Ecb{\Theta}{2}+
            \Ec{2}{3} \Ecb{2}{3} \Ec{\Theta}{1}\Ecb{\Theta}{1}\lsp
            \right)\lsp \Uc{3}{3}^2\,,\nonumber\\
\TAOdd[\llsp\prime\prime\llsp]{2} &=\big(
            \Ec{1}{3} (\Ecb{1}{3} \Uc{2}{2}+
            		   \Ecb{2}{3} \Uc{2}{1})+
            \Ec{2}{3} (\Ecb{1}{3} \Uc{1}{2}+
                       \Ecb{2}{3} \Uc{1}{1})\lnsp
            \big)\lsp \Uc{\Theta}{\Theta} \Uc{3}{3}^2\,,\nonumber\\
\TAOdd[\llsp\prime\prime\llsp]{3} &= \left(\Uc{1}{1} \Ec{\Theta}{2}\Ecb{\Theta}{2} +
                   \Uc{1}{2} \Ec{\Theta}{2}\Ecb{\Theta}{1} +
                   \Uc{2}{1} \Ec{\Theta}{1}\Ecb{\Theta}{2} +
                   \Uc{2}{2} \Ec{\Theta}{1}\Ecb{\Theta}{1}
\lsp\right)\lsp \Uc{3}{3}^3 \,,\nonumber\\
\TAOdd[\llsp\prime\prime\llsp]{4} &=\left(\Uc{1}{1} \Ec{2}{3}\Ecb{2}{3} +
                   \Uc{1}{2} \Ec{2}{3}\Ecb{1}{3} +
                   \Uc{2}{1} \Ec{1}{3}\Ecb{2}{3} +
                   \Uc{2}{2} \Ec{1}{3}\Ecb{1}{3}
\lsp\right)\lsp \Ec{\Theta}{3}\Ecb{\Theta}{3} \Uc{3}{3} \,,\nonumber\\
\TAOdd[\llsp\prime\prime\llsp]{5} &= i\lsp\left(
(\Uc{1}{3} \Ec{\Theta}{2} + \Uc{2}{3} \Ec{\Theta}{1}\lsp) \Ecb{\Theta}{3} \Uc{3}{1} \Uc{3}{2}
-
\Ec{\Theta}{3} (\Ecb{\Theta}{1}\Uc{3}{2}  +\Ecb{\Theta}{2} \Uc{3}{1} \lsp)  \Uc{1}{3} \Uc{2}{3}
\lsp\right)\lsp \Uc{3}{3} \,,\nonumber\\
\TAOdd[\llsp\prime\prime\llsp]{6} &=i\lsp\big(
(\Ec{1}{3}\Uc{2}{3} + \Ec{2}{3}\Uc{1}{3}\lsp)(\Uc{\Theta}{1}\Ecb{\Theta}{2}+\Uc{\Theta}{2}\Ecb{\Theta}{1}\lsp)\nonumber\\&\hspace{.42cm}
-
(\Ecb{1}{3}\Uc{3}{2} + \Ecb{2}{3}\Uc{3}{1}\lsp)(\Ec{\Theta}{1}\Uc{2}{\Theta}+\Ec{\Theta}{2}\Uc{1}{\Theta}\lsp)
\lsp\big)\lsp \Uc{3}{3}^2
\,.
\end{align}
\vspace{-\baselineskip}\vspace{8pt}
\newcommand{\AOddStructures}{\eqref{AOddStructures}\xspace}\\
If $\ell = 1$ the structures $\TAOdd[\llsp\prime\llsp]{1},\,\TAOdd[\llsp\prime\prime\llsp]{4}$ and $\TAOdd[\llsp\prime\prime\llsp]{5}$ are not present.

Next let us consider case \ref{caseB} for $\ell$ even. In \tBEvenSpin we have the following tensor structures:
\begin{align}\label{BEvenStructures}
\TBEven{1}&= \Ec{1}{3}\Ec{2}{3}(\Ecb{2}{3}\Uc{3}{1} +
              \Ecb{1}{3}\Uc{3}{2}\lsp)\,, \nonumber\\
\TBEven{2}&= \left(
                 \Ec{1}{3}(\Uc{2}{1}\Uc{3}{2}+\Uc{3}{1}\Uc{2}{2}\lsp)
                 +\Ec{2}{3}(\Uc{1}{1}\Uc{3}{1}+\Uc{3}{1}\Uc{1}{2}\lsp)
\lsp\right)\lsp \Uc{3}{3} \,, \nonumber\\
\TBEven[\llsp\prime\llsp]{1}&= \Ecb{1}{2} \left(\Ec{1}{3} \Ec{\Theta}{2} \Uc{3}{\Theta}+
                              \Ec{2}{3} \Ec{\Theta}{1} \Uc{3}{\Theta}\lsp\right)\lsp \Uc{3}{3}\,, \nonumber\\
\TBEven[\llsp\prime\llsp]{2}&= \Ecb{1}{2} \left(\Ec{1}{3} \Ec{\Theta}{3} \Uc{2}{\Theta}+
                              \Ec{2}{3} \Ec{\Theta}{3} \Uc{1}{\Theta}\lsp\right)\lsp \Uc{3}{3}\,, \nonumber\\
\TBEven[\llsp\prime\llsp]{3}&= \Ec{1}{2} \Uc{3}{1} \Uc{3}{2} \Ec{\Theta}{3}\Ecb{\Theta}{3} \,, \nonumber\\
\TBEven[\llsp\prime\llsp]{4}&= \Ec{1}{3} \Ec{2}{3} \Ecb{1}{2} \Ec{\Theta}{3}\Ecb{\Theta}{3} \,, \nonumber\\
\TBEven[\llsp\prime\llsp]{5}&= \Ec{1}{2} \left(\Uc{3}{2} \Ec{\Theta}{3}\Ecb{\Theta}{1}+
                              \Uc{3}{1} \Ec{\Theta}{3}\Ecb{\Theta}{2}\lsp\right)\lsp \Uc{3}{3}\,, \nonumber\\
\TBEven[\llsp\prime\llsp]{6}&= \Ec{1}{2} \left(\Uc{3}{2} \Uc{\Theta}{1}\Uc{3}{\Theta}+
                              \Uc{3}{1} \Uc{\Theta}{2}\Uc{3}{\Theta}\lsp\right)\lsp \Uc{3}{3}\,.
\end{align}
\vspace{-\baselineskip}\vspace{8pt}
\newcommand{\BEvenStructures}{\eqref{BEvenStructures}\xspace}\\
If $\ell = 0$ the structures $\TBEven{1},\,\TBEven[\llsp\prime\llsp]{3}$ and $\TBEven[\llsp\prime\llsp]{4}$ are not present.
The structures for the odd-$\ell$ case appearing in \tBOddSpin are
\eqna{
\TBOdd{1}&= \Ec{1}{3}\Ecb{1}{2}\Ec{2}{3}\Uc{3}{3}^2\,,\\
\TBOdd{2}&= \Ec{1}{2}\Uc{3}{1}\Uc{3}{2}\Uc{3}{3}^2\,,\\
\TBOdd[\llsp\prime\llsp]{1}&= \Ec{1}{3}\Ec{2}{3}\left(
\Ecb{2}{3} \Uc{3}{1}+\Ecb{1}{3} \Uc{3}{2}
\lsp\right)\lsp\Ec{\Theta}{3}\Ecb{\Theta}{3}\,, \\
\TBOdd[\llsp\prime\llsp]{2}&= \left(
(\Uc{1}{1} \Uc{3}{2} + \Uc{3}{1} \Uc{1}{2}\lsp)\Ec{\Theta}{3} \Uc{2}{\Theta}
+
(\Uc{2}{1} \Uc{3}{2} + \Uc{3}{1} \Uc{2}{2}\lsp)\Ec{\Theta}{3} \Uc{1}{\Theta}
\lsp\right)\lsp\Uc{3}{3}^2\,, \\
\TBOdd[\llsp\prime\llsp]{3}&= \left(
\Ec{2}{3}\lsp(\Uc{3}{2} \Ec{\Theta}{1} \Ecb{\Theta}{1} + \Uc{3}{1} \Ec{\Theta}{1} \Ecb{\Theta}{2}\lsp)
-
\Ec{1}{3}\lsp(\Uc{3}{2} \Ec{\Theta}{2} \Ecb{\Theta}{1} + \Uc{3}{1} \Ec{\Theta}{2} \Ecb{\Theta}{2}\lsp)
\lsp\right)\lsp\Uc{3}{3}^2\,, \\
\TBOdd[\llsp\prime\llsp]{4}&= \Ec{1}{3}\Ec{2}{3}\left(
\Ecb{1}{3} \Ec{\Theta}{3}\Ecb{\Theta}{2}
+\Ecb{2}{3} \Ec{\Theta}{3}\Ecb{\Theta}{1}
\lsp\right)\lsp\Uc{3}{3}\,, \\
\TBOdd[\llsp\prime\llsp]{5}&=  \Uc{3}{1}\Uc{3}{2}\left(
\Ec{1}{3} \Ec{\Theta}{2}\Ecb{\Theta}{3}
+\Ec{2}{3} \Ec{\Theta}{1}\Ecb{\Theta}{3}
\lsp\right)\lsp\Uc{3}{3}\,, \\
\TBOdd[\llsp\prime\llsp]{6}&= \left(
\Ec{1}{3}(\Uc{2}{1}\Uc{3}{2} + \Uc{3}{1}\Uc{2}{2} \lsp)
+
\Ec{2}{3}(\Uc{1}{1}\Uc{3}{2} + \Uc{3}{1}\Uc{1}{2} \lsp)
\lsp\right)\lsp\Ec{\Theta}{3}\Ecb{\Theta}{3}\Uc{3}{3}\,, \\
\TBOdd[\llsp\prime\llsp]{7}&= \Uc{3}{1}\Uc{3}{2}\left(\Ec{\Theta}{1}\Uc{2}{\Theta}+
                                \Ec{\Theta}{2}\Uc{1}{\Theta}
                                \lsp\right)\lsp\Uc{3}{3}^2\,,\\
\TBOdd[\llsp\prime\llsp]{8}&= \left(
(\Ec{1}{3}\Uc{2}{2}+\Ec{2}{3}\Uc{1}{2}\lsp)\Ec{\Theta}{3}\Ecb{\Theta}{1}
-
(\Ec{1}{3}\Uc{2}{1}+\Ec{2}{3}\Uc{1}{1}\lsp)\Ec{\Theta}{3}\Ecb{\Theta}{2}
\lsp\right)\lsp\Uc{3}{3}^2\,, \\
\TBOdd[\llsp\prime\llsp]{9}&= \Ec{1}{2}\Ecb{1}{2}\Ec{\Theta}{3}\Uc{3}{\Theta}\Uc{3}{3}^2\,.
}[BOddStructures]
If $\ell = 1$ the structure $\TBOdd[\llsp\prime\llsp]{1}$ is not present.

Finally let us consider case \ref{caseC} for $\ell$ even. In \tCEvenSpin we have the following tensor structures:
\eqna{
\TCEven{1}&= \Ec{1}{3}\Ec{2}{3}\Uc{3}{1}\Uc{3}{2}
 \,, \\
\TCEven[\llsp\prime\llsp]{1}&= \Ec{1}{2} \Uc{3}{1}\Uc{3}{2} \Ec{\Theta}{3}\Uc{3}{\Theta}\,,\\
\TCEven[\llsp\prime\llsp]{2}&= \Ec{1}{3}\Ec{2}{3}\Ecb{1}{2}\Ec{\Theta}{3}\Uc{3}{\Theta}\,.
}[CEvenStructures]
The structures for the odd-$\ell$ case appearing in \tCOddSpin are
\eqna{
\TCOdd{1}&= \Ec{1}{3}\Ec{2}{3}\left(
             \Uc{3}{2}\Ec{\Theta}{3} \Ecb{\Theta}{1}
             +
             \Uc{3}{1}\Ec{\Theta}{3} \Ecb{\Theta}{2}
\lsp\right)\Uc{3}{3}\,, \\
\TCOdd{2}&= \Uc{3}{1}\Uc{3}{2}\left(
            \Ec{1}{3}\Ec{\Theta}{2} \Uc{3}{\Theta}
           +\Ec{2}{3}\Ec{\Theta}{1} \Uc{3}{\Theta}
\lsp\right)\Uc{3}{3}\,, \\
\TCOdd{3}&= \Uc{3}{1}\Uc{3}{2}\left(
            \Ec{1}{3}\Ec{\Theta}{3} \Uc{2}{\Theta}
           +\Ec{2}{3}\Ec{\Theta}{3} \Uc{1}{\Theta}
\lsp\right)\Uc{3}{3}\,, \\
\TCOdd{4}&= \Ec{1}{3}\Ec{2}{3}\Uc{3}{1}\Uc{3}{2}\Ec{\Theta}{3}\Ecb{\Theta}{3} \,.
}[COddStructures]

\newsec{Tensor structures in vector formalism}
\label{appendix:structureLorentz}

The structures in $t_{\text{A}}$ can be found by considering the problem
purely in Lorentz vector indices. For explicit computations it is of course
more convenient to use spinor indices, which we can contract with auxiliary
commuting spinors---this relieves us from the headache of removal of
traces. In vector form we have\foot{In these vector-form expressions a
subtraction of the traces in the indices $\rho_1,\ldots,\rho_\ell$ is
understood.}
\eqna{
t_{\text{A}\lsp\mu\nu;\lsp\rho_1\lnsp\ldots\rho_\ell}^
  {\lsp(\ell\text{ even})}(X,\Xb)&=
  \frac{1}{(X\cdot\Xb)^{3+\frac12\ell-q}} \left(\sum_{i=1}^{4} \TAEvenLorentz{i}_{\mu\nu;(\rho_1\rho_2} \left(\mathcal{A}_i +\mathcal{B}_i \lsp \beta^2\right) +  \sum_{i=5}^{12} \TAEvenLorentz{i}_{\mu\nu;(\rho_1\rho_2} \mathcal{C}_i\right) \lsp U_{\lnsp\rho_3}\cdots U_{\lnsp\rho_\ell)} \,,\\
    \beta^2& = \frac{V^2}{X\cdot\Xb}
  \,,}[tAEvenSpin]
where $\mathcal{A}_i,\,\mathcal{B}_i$ and $\mathcal{C}_i$ are real constants. The tensor structures $ \mathcal P^{(i)}$ are given by
\eqna{
  \begin{aligned}
 \TAEvenLorentz{1}_{\mu\nu;\rho_1\rho_2} &=  \lsp\eta_{\mu\nu}\llsp  U_{\lnsp\rho_1} U_{\lnsp\rho_2}\,,
 \qquad & \TAEvenLorentz{7}_{\mu\nu;\rho_1\rho_2} &=    V_{[\mu}\eta_{\nu]\rho_1} U_{\lnsp\rho_2}\,, \\
\TAEvenLorentz{2}_{\mu\nu;\rho_1\rho_2} &=\frac{X_{\lnsp(\mu}  \Xb_{\lnsp\nu)}}{X\cdot\Xb} U_{\lnsp\rho_1}U_{\lnsp\rho_2}\,,
\qquad &  \TAEvenLorentz{8}_{\mu\nu;\rho_1\rho_2} &= i\lsp\eta_{\rho_1[\mu}\epsilon_{\nu]\kappa\lambda\rho_2}  X^{\kappa} \Xb^{\lambda} \,,\\
  \TAEvenLorentz{3}_{\mu\nu;\rho_1\rho_2} &= U_{\lnsp(\mu}  \eta_{\nu)\rho_1} U_{\lnsp\rho_2}\,,
\qquad & \TAEvenLorentz{9}_{\mu\nu;\rho_1\rho_2} &=   i\frac{\lsp U\lnsp\cdot\lnsp V}{X\cdot\Xb} \lsp \epsilon_{\mu\nu\kappa\rho_1}\lsp U^\kappa\lsp  U_{\lnsp\rho_2}\,,\\
 \TAEvenLorentz{4}_{\mu\nu;\rho_1\rho_2} &=X\cdot\Xb\lsp \eta_{\rho_1(\mu}\eta_{\nu)\rho_2}\,,
 \qquad  &\TAEvenLorentz{10}_{\mu\nu;\rho_1\rho_2}& = i\lsp  \epsilon_{\mu\nu\kappa\rho_1}\lsp U^\kappa\lsp V_{\rho_2}\,,\\
 \TAEvenLorentz{5}_{\mu\nu;\rho_1\rho_2} &= \frac{ i\lsp \epsilon_{\mu\nu\kappa\lambda} X^{\kappa}  \Xb^{\lambda}}{{X\cdot\Xb}} U_{\lnsp\rho_1} U_{\lnsp\rho_2}\,,
 \qquad  &\TAEvenLorentz{11}_{\mu\nu;\rho_1\rho_2}& =  \frac{\lsp\eta_{\rho_1[\mu}\lsp U_{\nu]} \lsp U\lnsp\cdot\lnsp V \lsp U_{\lnsp\rho_2}}{X\cdot\Xb}\,, \\
\TAEvenLorentz{6}_{\mu\nu;\rho_1\rho_2} &=  \epsilon_{\mu\nu\lambda\rho_1}V^{\lambda}\lsp U_{\lnsp\rho_2}\,,
 \qquad  &\TAEvenLorentz{12}_{\mu\nu;\rho_1\rho_2}& = i\lsp U_{[\mu}\epsilon_{\nu]\kappa\lambda\rho_1}\frac{X^{\kappa}  \Xb^{\lambda}}{X\cdot\Xb} U_{\lnsp\rho_2}\,,
 \end{aligned}
}[]
with the definitions\foot{$U$ and $V$ are called $Q$ and $P$, respectively,
in~\cite{Osborn:1998qu}.}
\eqn{U=\tfrac12(X+\Xb)\,,\qquad V=i(X-\Xb)\,.}[UandVdef]
The structures $\mathcal P^{(1)},\,\mathcal P^{(2)},\,\mathcal P^{(5)}$ in \tAEvenSpin appear for
$\ell=0$~\cite{Osborn:1998qu}, while the others do not. Since $V\to0$ and
$\Xb\to X$ if the Grassmann variables are set to zero, we have recovered
the four structures expected from the results of~\cite{Echeverri:2015rwa},
associated with the coefficients $\mathcal{A}_1,\,\mathcal{A}_2,\,\mathcal{A}_3$ and $\mathcal{A}_4$.

In the case of odd spins and before application of the Ward identity for
the supercurrent there is a parity-odd and a parity-even structure if all
Grassmann variables are set to zero~\cite{Echeverri:2015rwa}.
Additionally, there are nilpotent structures.  We find
\eqna{
  t_{\text{A}\lsp\mu\nu;\lsp\rho_1\lnsp\ldots\rho_\ell}^
  {\lsp(\ell\text{ odd})}(X,\Xb)&= \frac{1}{(X\cdot \Xb)^{3+\frac{1}{2}\ell-q}}\Bigg(\sum_{i=1}^2 \TAOddLorentz{i}_{\mu\nu;(\rho_1\rho_2\rho_3}\left(\mathcal{A}_i+\mathcal{B}_i \lsp \beta^2\right)+\sum_{i=3}^6 \TAOddLorentz[\lambda]{i}_{\mu\nu;(\rho_1\rho_2\rho_3}\left(\mathcal{C}_i\lsp V_\lambda+\mathcal{D}_i \lsp \zeta_\lambda\right)\\
  &\hspace{3.25cm}
  +\sum_{i=7}^{10}\TAOddLorentz{i}_{\mu\nu;(\rho_1\rho_2\rho_3}\lsp \mathcal{E}_i\Bigg)\lsp U_{\lnsp\rho_4}\cdots  U_{\lnsp\rho_{\ell})}\,,
  \\
      \beta^2& = \frac{V^2}{X\cdot\Xb}\,,\qquad\zeta_\mu = \frac{U\cdot V}{X\cdot \Xb}\lsp U_\mu
  \,,
}[tAOddSpin]
where $\mathcal{A}_i,\,\mathcal{B}_i,\,\mathcal{C}_i,\,\mathcal{D}_i,\,\mathcal{E}_i$ are real constants and the tensor structures $ \TAOddLorentz{i}$ are given by
\eqna{\begin{aligned}
\TAOddLorentz{1}_{\mu\nu;\rho_1\rho_2\rho_3} &=  \epsilon_{\mu\nu\lambda\rho_1}\lsp U^\lambda U_{\lnsp\rho_2} U_{\lnsp\rho_3}\,,
 \qquad & \TAOddLorentz[\lambda]{6}_{\mu\nu;\rho_1\rho_2\rho_3} &= X\cdot
 \Xb\lsp\eta_{\rho_1(\mu}\eta_{\nu)\rho_2}\lsp \delta^\lambda_{\,\,\rho_3}\,,\\
\TAOddLorentz{2}_{\mu\nu;\rho_1\rho_2\rho_3} &=U_{[\mu} \eta_{\nu](\rho_1}\lnsp U_{\lnsp\rho_2} U_{\lnsp\rho_3}\,,
\qquad &  \TAOddLorentz{7}_{\mu\nu;\rho_1\rho_2\rho_3} &= \frac{U_{\lnsp(\mu }V_{\lnsp\nu)}U_{\lnsp \rho_1}U_{\lnsp \rho_2}U_{\lnsp \rho_3}}{X\cdot \Xb}\,, \\
\TAOddLorentz[\lambda]{3}_{\mu\nu;\rho_1\rho_2\rho_3} &=
\eta_{\mu\nu}\,\delta^\lambda_{\,\,\rho_1}
U_{\lnsp\rho_2} U_{\lnsp\rho_3} \,,
\qquad & \TAOddLorentz{8}_{\mu\nu;\rho_1\rho_2\rho_3} &= \eta_{\rho_1(\mu}\epsilon_{\nu)\kappa\lambda\rho_2}X^\kappa \Xb^\lambda U_{\lnsp\rho_3}\,,\\
\TAOddLorentz[\lambda]{4}_{\mu\nu;\rho_1\rho_2\rho_3}
&=\delta^\lambda_{\,\,\lnsp(\mu} \,\eta_{\nu)\rho_1} U_{\lnsp\rho_2} U_{\lnsp\rho_3}\,,
 \qquad  &\TAOddLorentz{9}_{\mu\nu;\rho_1\rho_2\rho_3}& =\lsp U_{(\mu} \eta_{\nu)\rho_1} V_{\rho_2}
  U_{\lnsp\rho_3}\,,\\
\TAOddLorentz[\lambda]{5}_{\mu\nu;\rho_1\rho_2\rho_3}
&=\frac{\delta^\lambda_{\,\,\rho_1}  \lsp U_{\lnsp\mu}U_{\lnsp\nu}U_{\lnsp\rho_2}U_{\lnsp\rho_3}}{X\cdot \Xb}\,,
 \qquad  &\TAOddLorentz{10}_{\mu\nu;\rho_1\rho_2\rho_3}& = \frac{U_{(\mu}\epsilon_{\nu)\kappa\lambda\rho_1}\lsp
  X^{\kappa}\Xb^{\lambda}\lsp
  U_{\lnsp\rho_2}  U_{\lnsp\rho_3}}{X\cdot \Xb}\,.\\
 \end{aligned}
}[]
When all Grassmann variables are set to zero the parity-odd structure
associated with $\mathcal{A}_1$ and the parity-even structure associated
with $\mathcal{A}_2$ are the only ones that survive.

\newsec{Special cases in the solutions of the Ward identities}
\label{appendix:WISpinZeroOne}
In Sec.~\ref{Ward} we studied the Ward identities arising from equation
\SupercurCons assuming generic values for $\Delta,\ell$. As one can see by
inspecting equations \WardtAEvenSol, \WardtAOddSol, \WardtBEvenSol,
\WardtBOddSol, \WardtCEvenSol and \WardtCOddSol all terms are well defined
for unitary values of $\Delta$, except for the special cases $\ell = 0$ in
\WardtAEvenSol and $\ell =1$ in \WardtAOddSol for which we find a pole at
$\Delta = \ell + 2$, i.e.\ the unitarity bound. These poles are artifacts
of the choice of independent coefficients made when solving the equations
arising from conservation and can therefore be removed. A choice which is
valid for all unitary values of $\Delta$ and $\ell$ does not exist.

Case \ref{caseA} for $\ell = 0$ has only the structures associated with the
coefficients $\AevenA$, $\AevenB$, $\AevenC$, $\AevenD$ and $\AevenE$. If
we choose to solve for $\AevenC$ we obtain simply
\eqn{
\AevenA=\AevenB=\AevenD=\AevenE=0\,.
}[WardtAEvenZero]
Since we are at the unitarity threshold most of the coefficients of the expansion in $\theta_3\thetab_3$ computed in Sec.~\ref{ThetaThreeExpansion} vanish. We are left with only
\eqn{
\lambda_{++}^{(-)} = 2 i \lsp \AevenC\,.
}[]
\par
Case \ref{caseA} for $\Delta=3, \ell = 1$ corresponds to the third operator
being the Ferrara--Zumino multiplet $\CJ_{\gamma\gammad}$. The structures
associated with $\AoddM,\,\AoddN,\,\AoddL$ and $\AoddO$ are not present. Solving for $\AoddA$ and $\AoddJ$ we find
\eqn{\begin{aligned}
\AoddC&=\AoddB=\AoddD=\AoddP=0\,,\\
\AoddI&=-\tfrac32 \AoddK=\AoddJ\,,&\quad
\AoddE&=-\tfrac12\AoddF=2 \AoddA
\\
\AoddG&=-2 \AoddA-\tfrac13 \AoddJ\,,&\quad\AoddH&=4 \AoddA+\AoddJ\,.
\end{aligned}}[]
The coefficients of the expansion in $\theta_3\thetab_3$ are
\eqn{
\lambda_{++}^{(1)} = -2 \AoddA - \tfrac16 \AoddJ\,,\qquad
\lambda_{++}^{(2)} = -\lambda_{++}^{(3)} = \tfrac23(3\AoddA+ \AoddJ)\,,\qquad
\lambda_{++}^{(4)} = -\tfrac13 \AoddJ\,,\qquad
\lambda_{++}^{(5)} = \tfrac12(4 \AoddA + \AoddJ)\,,
}[]
with all others being zero. This case was already analyzed
in~\cite{Osborn:1998qu} with the same results. The relation between our
coefficients and the coefficients $A,\ldots,J$ defined in
\cite{Osborn:1998qu} is
\eqn{\begin{aligned}
A &= \AoddA\,,& D &= \tfrac{1}{4}(\AoddF - \AoddJ - \AoddH -\AoddK) \,,  & G &= \tfrac{1}{2}\AoddI\,,\\
B &= \tfrac{1}{8}(4 \lsp\AoddA - \AoddB)\,, & E &= \tfrac{1}{2}\AoddG \,,& H &= \AoddJ\,,\\
C &= \tfrac{1}{8}(\AoddE-\AoddI-\AoddG)\,, & F &= \AoddH\,, & J &= \tfrac{1}{2}\AoddK\,.\\
\end{aligned}
}[OsbornOur]

Taking the spin-two superconformal descendant of each superconformal
primary yields the three-point correlator of the stress-energy tensor,
$T_{\mu\nu}$. It is interesting, thus, to relate the coefficients defined
here, $\AoddA$ and $\AoddJ$, with the anomaly coefficients $c$
(proportional to the central charge $C_T$) and $a$ (Euler anomaly).
Using~\cite[Eq.\ (11.7)]{Osborn:1998qu} together with~\OsbornOur, or,
equivalently, following~\cite[Appendix C]{Hofman:2016awc} and using the
relation between $C_J$ and $C_T$ stemming from supersymmetry~\cite[Eq.\
(5.5)]{Li:2014gpa} one obtains
\eqn{
\AoddA = -\frac{8}{9\pi^6}(5a-3c)\,,\qquad \AoddJ = \frac{16}{3\pi^6}(2a-3c)
\,.
}[]
The precise relation between $c$ and $C_T$ is~\cite{Hofman:2016awc} $c =\frac{\pi^4}{40}C_T$.

\newsec{Conventions for the supersymmetric derivatives}
\label{Conventions}

In order to impose \SupercurCons we follow the formalism of~\cite{Osborn:1998qu} and pass the derivatives $D_{1\alpha}$ and $\Db_{1\alphad}$ through the prefactor in \ThreePFAgen. This can be done using the identities
\eqna{
{\Db_1}^\alphad \left(\frac{(\xup_{3\bar1})_{\alpha\alphad}}{({x_{\bar1 3}}^2)^2}F^\alpha(X,\Theta,\Thetab)\right) &= -i \frac{1}{{x_{\bar3 1}}^2{x_{\bar1 3}}^2}\mathcal{D}_{\alpha} F^\alpha(X,\Theta,\Thetab)\,,\\
{D_1}^\alpha \left(\frac{(\xup_{1\bar3})_{\alpha\alphad}}{({x_{\bar3 1}}^2)^2}F^\alphad(X,\Theta,\Thetab)\right) &= -i \frac{1}{{x_{\bar3 1}}^2{x_{\bar1 3}}^2}\overbar{\mathcal{D}}_{\alphad} F^\alphad(X,\Theta,\Thetab)\,.
}[]
The superspace derivatives are defined in the following way
\eqn{
D_{i\alpha} = \frac{\partial}{\partial \theta_i^\alpha} + i  \sigma^\mu_{\alpha\alphad}\lsp\thetab_i^\alphad \frac{\partial}{\partial x_i^\mu}\,,\qquad
\Db_{i\alphad} = -\frac{\partial}{\partial \thetab_i^\alphad} - i  \theta_i^\alpha\sigma^\mu_{\alpha\alphad}\frac{\partial}{\partial x_i^\mu}\,.
}[]
For the derivatives $\mathcal{D}$ and $\overbar{\mathcal{D}}$ we can use three different representations according to whether we prefer writing the expressions in terms of $X,\,\Xb$ or $U = \frac12(X+\Xb)$.
\begin{subequations}
\begin{align}
\label{OsbornDefCDCDb}
\mathcal{D}_{\alpha} &= \frac{\partial}{\partial \Theta^\alpha} - 2 i  \sigma^\mu_{\alpha\alphad}\lsp\Thetab^\alphad \frac{\partial}{\partial X^\mu}\,,&\qquad
\overbar{\mathcal{D}}_{\alphad} &= -\frac{\partial}{\partial \Thetab^\alphad}\,,
\\\label{OsbornDefCDCDb2}
\mathcal{D}_{\alpha} &= \frac{\partial}{\partial \Theta^\alpha}\,,&\qquad
\overbar{\mathcal{D}}_{\alphad} &= -\frac{\partial}{\partial \Thetab^\alphad}+2 i  \Theta^\alpha\sigma^\mu_{\alpha\alphad} \frac{\partial}{\partial \Xb^\mu}\,,
\\\label{OsbornDefCDCDbU}
\mathcal{D}_{\alpha} &= \frac{\partial}{\partial \Theta^\alpha} -  i  \sigma^\mu_{\alpha\alphad}\lsp\Thetab^\alphad \frac{\partial}{\partial U^\mu}\,,&\qquad
\overbar{\mathcal{D}}_{\alphad} &= \frac{\partial}{\partial \Thetab^\alphad}+ i  \Theta^\alpha\sigma^\mu_{\alpha\alphad} \frac{\partial}{\partial U^\mu}\,.
\end{align}
\end{subequations}

\end{appendices}

\bibliography{R-current_3PFs_in_4d_N=1_SCFTs}

\end{document}